\documentclass[11pt,a4paper]{article}
\pdfoutput=1
\usepackage{jcappub}
\usepackage{graphicx}
\usepackage{caption}
\usepackage{placeins}
\usepackage{subcaption}
\usepackage{amsmath,bm}
\usepackage[T1]{fontenc}
\usepackage{rotating}
\usepackage{float}
\usepackage[normalem]{ulem}
\usepackage{mathrsfs}

\floatstyle{plaintop}
\restylefloat{table}

\definecolor{darkgreen}{RGB}{0,170,0}
\definecolor{darkgray}{RGB}{80,80,80}

\newcommand{\kmps}{\mbox{km}/\mbox{s}}
\newcommand{\threej}[6]{ \begin{pmatrix}
   #1 & #2 & #3 \\
   #4 & #5 & #6 
  \end{pmatrix}}

\begin{document}

\title{Inelastic dark matter nucleus scattering}

\author[a,b,c]{G. Arcadi,}
\emailAdd{giorgio.arcadi@uniroma3.it}
\author[b]{C. D\"oring,}
\emailAdd{christian.doering@mpi-hd.mpg.de}
\author[b]{C. Hasterok}
\emailAdd{constanze.hasterok@mpi-hd.mpg.de}
\author[b]{and S. Vogl}
\emailAdd{stefan.vogl@mpi-hd.mpg.de}
\affiliation[a]{Dipartimento di Matematica e Fisica, Universit\`a di Roma
3, Via della Vasca Navale 84, 00146, Roma, Italy.}
\affiliation[b]{Max-Planck-Institut f\"ur Kernphysik, Saupfercheckweg 1, D-69117 Heidelberg, Germany}
\affiliation[c]{INFN Sezione Roma 3}

\abstract{
Direct detection experiments aim at the detection of dark matter in the form of weakly interacting massive particles (WIMPs) by searching for signals from elastic dark matter nucleus scattering. Additionally, inelastic scattering in which the nucleus is excited is expected from nuclear physics and provides an additional detectable signal. In the context of a low-energy effective field theory we investigate the experimental reach to these inelastic transitions for xenon-based detectors employing a dual-phase time projection chamber. We find that once a dark matter signal is established, inelastic transitions enhance the discovery reach and we show that they allow a better determination of the underlying particle physics.}

\maketitle

\section{Introduction}\label{sec:intro}

The search for dark matter (DM) in the form of weakly interacting massive particles (WIMPs) with direct detection experiments has made great progress in recent years \cite{Aprile:2013doa,Xe1T-SI,LUX-SI,PandaX-SI}. With the next generation of experiments on the way it is a timely question to ask what we can expect to learn about the properties of WIMPs once a signal has been established. We do know neither the mass of the DM particle nor other properties such as the spin. In addition, the interactions between WIMPs and Standard Model (SM) particles are not understood either. We hope to be able to shed more light on these parameters once a detection is established. In the following we will use the terms DM and WIMP interchangeably.

 Direct detection aims to observe the recoil induced by DM scattering off nuclei. The typical velocity of DM bound to the Milky Way is limited by the galactic escape velocity $v_{esc}\approx550 \,\kmps$ and, consequently, the highest nuclear recoil energy for an earth-based target assuming WIMPs with a mass of $100 \,\mbox{GeV}$ is $\approx 300\;\mbox{keV}$. Therefore, direct detection experiments are testing the non-relativistic limit of the DM nucleus interaction and it is possible to describe the scattering process in terms of a non-relativistic effective theory of DM and nucleons. The general operator basis for non-relativistic DM-nucleon interactions and the formalism for the description of DM scattering off nuclei has been developed by \cite{Fitzpatrick:2012ix}. The non-relativistic effective theory can be used for various applications~\cite{DelNobile:2013sia,Fitzpatrick:2012ib,Catena:2014uqa,Catena:2015uha,Catena:2016hoj,Dent:2015zpa,Gresham:2014vja,Boddy:2018kfv,Kavanagh:2015jma}. Experimental constraints on the effective operators from dedicated analyses are reported in~\cite{Aprile:2017aas,Angloher:2018fcs}.

Different combinations of WIMP mass and WIMP-nucleus interactions can lead to very similar signals in direct detection experiments. Therefore, a large exposure or additional signals are needed to gain information on the properties of DM. Typically, a large exposure is achieved with bigger and more sensitive detectors than available to date and is therefore very costly. However, direct detection experiments are sensitive to inelastic WIMP-nucleus scattering in addition to the standard signature of elastic WIMP nucleus scattering \cite{Goodman:1984dc,Ellis:1988nb,Engel:1999kv}.
In this process, the nucleus is excited and due to the different nuclear structure of the ground and the excited state the response depends crucially on the DM-nucleon interaction.
Exploiting the inelastic signal enhances the experimental sensitivity to DM properties without necessitating any modifications on the detector setup.  This possibility is particularly attractive in the context of direct detection experiments employing dual-phase time projection chambers (TPC) with liquid xenon (LXe) targets \cite{Xe1T-Exp,LUX-Exp,PandaX-Exp}. In addition to being the leading technology for the standard channel of direct WIMP searches \cite{Xe1T-SI,LUX-SI,PandaX-SI}, these detectors are well suited for inelastic searches \cite{Aprile:2017ngb,PandaX-Inelastic,Suzuki:2018xek} since two common isotopes, $^{129}$Xe and $^{131}$Xe, have unusually low-energetic exited states with an excitation energy of $39.6 \,\mbox{keV}$ and $80.2 \,\mbox{keV}$, respectively.
 It is well known that inelastic scattering can help to distinguish the spin-dependent (SD) and spin-independent (SI) interactions  traditionally considered in direct detection~\cite{Ellis:1988nb}.   A detailed  nuclear physics calculation of the inelastic SD DM nucleon scattering in xenon has been presented in \cite{Baudis:2013bba} and  the sensitivity of a XENONnT-like experiment with a  $15\, \mbox{tonne} \times \mbox{year}$ exposure has been studied  in \cite{McCabe:2015eia}. The absence of signals of this type can be re-interpreted as an upper bound on conventional SD interactions with the strongest being of $\sim 10^{-38}\,{\mbox{GeV}}^{-2}$ for a WIMP mass of 200 GeV~\cite{Suzuki:2018xek}.
 
The aim of this work is three fold: First,
we include in our study the full set of non-relativistic effective operators introduced
in \cite{Fitzpatrick:2012ix} rather then just the conventional spin-independent and spin-dependent interactions.
Second, we analyze the discovery reach to these operators for future experiments in light of recent null results and derive the prospects for an ultimate future direct detection experiment such as DARWIN~\cite{DARWIN}. Third, we investigate the impact of inelastic scattering on our ability to pin down the DM-nucleon interaction once a signal has been detected. 
 
 The structure of this paper is as follows: We briefly introduce the non-relativistic effective DM-nucleon theory  and present the modifications required to describe inelastic scattering in Sec.~\ref{sec:inelTheory}. Next, we discuss the general properties of the signal and  outline the process of signal creation and detection in LXe TPCs in Sec.~\ref{sec:xenonExp}. The detector model employed for the following studies is described in detail. In Sec.~\ref{sec:DiscoveryReach} the prospects for discovering an inelastic signal with future experiments are discussed while Sec.~\ref{sec:DiscriminationPower} focuses on the use of the additional information gained from an analysis of inelastic scattering for the determination of the DM-nucleon interaction. Finally, we present our conclusions in Sec.~\ref{sec:Conclusions}. Additional information regarding the description of inelastic scattering and the detector model are presented in the Appendix.

\section{Inelastic scattering in non-relativistic effective theory}
\label{sec:inelTheory}
In our analysis we use the parametrization of DM-nucleon interactions in terms of  non-relativistic effective operators~\cite{Fitzpatrick:2012ix}\footnote{The operators entering at the nucleon-DM level can be mapped to the operators of a theory of relativistic DM \cite{DelNobile:2013sia,Bishara:2017pfq,Hoferichter:2018acd}. Since we are interested in low-energy observables we prefer to remain agnostic regarding the UV-completion.}. In this section we briefly review the formalism for calculating the DM nucleus scattering rates and comment on the extensions that are required to incorporate inelastic scattering. For a detailed introduction to the non-relativistic effective field theory of DM and nucleons we refer to  \cite{Fitzpatrick:2012ix,Anand:2013yka}. A more concise summary can be found for example in \cite{Catena:2015uha}. A Mathematica package which facilitates the computation of the different cross sections for elastic scattering has been published \cite{Anand:2013yka,dmformfactor}. We base our implementation of inelastic scattering on this package and extend it where necessary.

In the non-relativistic limit relevant for direct detection experiments, the form of the interaction between DM and nucleons is constrained by energy and momentum conservation and by Galilean invariance, i.e. invariance with respect to shifts in
the three-dimensional particle velocity. Interactions respecting these constraints can be described by combinations of five operators, $1,$ $\vec{S}_\chi$, $\vec{S}_N$, $\vec{v}^\perp=\vec{v}+\frac{\vec{q}}{2\mu_N}$, and $i\vec{q}$, which act on the two-particle Hilbert space spanned by the nucleon and DM states.
In the following, we limit ourselves to fermionic DM and  to operators that are at most linear in $\vec{S}_\chi$, $\vec{S}_N$ and $\vec{v}^\perp$. With these constraints 14 linearly independent operators can be constructed\footnote{ Note that a further operator, $\mathcal{O}_2=\vec{v}^\perp \cdot \vec{v}^\perp$, cannot arise as the leading operator from a complete UV-theory and has therefore been omitted in \cite{Anand:2013yka}. For spin-1 DM four additional operators can be generated \cite{Dent:2015zpa,Catena:2019hzw}. We will not consider them here but they can be straightforwardly included in an analysis of inelastic scattering.}, see Tab.~\ref{tab:Operators}.
\begin{table}
\centering
\begin{tabular}{l l}
\hline 
$\mathcal{O}_1= 1_\chi 1_N$    &  $\mathcal{O}_3=i \vec{S}_N \cdot \left(\frac{\vec{q}}{m_N}\times \vec{v}^\perp \right) 1_\chi$ \\
$ \mathcal{O}_4=\vec{S}_\chi \cdot \vec{S}_N$      &  $\mathcal{O}_5=i \vec{S}_\chi \cdot \left(\frac{\vec{q}}{m_N} \times \vec{v}^\perp\right) 1_N $ \\
$\mathcal{O}_6=\left(\vec{S}_\chi \cdot \frac{\vec{q}}{m_N}\right)\left(\vec{S}_N \cdot \frac{q}{m_N}\right)$ & $\mathcal{O}_7=\vec{S}_N \cdot \vec{v}^\perp 1_\chi$ \\
$\mathcal{O}_8=\vec{S}_\chi \cdot \vec{v}^\perp 1_N $ & $\mathcal{O}_9=i \vec{S}_\chi \cdot \left(\vec{S}_N \times \frac{\vec{q}}{m_N}\right)$ \\
$\mathcal{O}_{10}=i \vec{S}_N \cdot \frac{\vec{q}}{m_N} 1_\chi$ & $\mathcal{O}_{11}=i \vec{S}_\chi \cdot \frac{\vec{q}}{m_N} 1_N$ \\
$\mathcal{O}_{12}=\vec{S}_\chi \cdot \left(\vec{S}_N \times \vec{v}^\perp\right)$ & $ \mathcal{O}_{13}=i \left(\vec{S}_\chi \cdot \vec{v}^\perp\right) \left(\vec{S}_N \cdot \frac{\vec{q}}{m_N}\right)$ \\
$\mathcal{O}_{14}=i \left(\vec{S}_\chi \cdot \frac{\vec{q}}{m_N}\right)\left(\vec{S}_N \cdot \vec{v}^\perp\right)$ & $\mathcal{O}_{15}=-\left(\vec{S}_\chi \cdot \frac{\vec{q}}{m_N}\right)\left(\left(\vec{S}_N \times \vec{v}^\perp\right)\cdot \frac{\vec{q}}{m_N}\right)$ \\
\hline
\end{tabular}
\caption{\label{tab:Operators} Non-relativistic effective operators for DM-nucleon
interactions. We follow the numbering scheme of~\cite{Anand:2013yka}. }
\end{table}
Linear combinations of these operators form the Hamiltonian density
\begin{equation}
\mathcal{H}=\sum_{\tau=0,1} \sum_{k=1}^{15} c_k^{\tau} \mathcal{O}_k t^\tau,
\label{eq:HamDMnucleon}
\end{equation}
where the operators $t^\tau$ act on isospin states with $t^0= 1$ and $t^1=\tau_3$ (third Pauli matrix). The coefficients  $c_k^{t}$ parametrize the strength of the isoscalar and isovector interactions and can be recast in terms of the interaction strength with protons and neutrons via $c_k^p=(c^0_k +c^1_k)/2$ and $c_k^n=(c^0_k -c^1_k)/2$. Note that $c_k^i$ have dimension of mass$^{-2}$ due to the normalization convention chosen in \cite{Anand:2013yka}. Consequently, we are implicitly encoding in the $c_k^{p,n}$ coefficients a New Physics scale $\Lambda$ such that $c_k^i= \frac{1}{\Lambda^2}$. 



The differential recoil cross section for DM scattering off a target nucleus of mass $m_T$ and spin $J$ is given by 
\begin{align}
\frac{\mbox{d}\sigma_T(v^2,E_R)}{\mbox{d} E_R}= \frac{m_T}{2\pi v^2} \langle |\mathcal{M}_{\rm eff}|^2 \rangle
\end{align} 
where $ \langle |\mathcal{M}_{\rm eff}|^2 \rangle$ denotes the spin-summed and spin-averaged square matrix element of DM-nucleus scattering and $v$ is the speed of the incoming DM particle in the rest frame of the target nucleus. 
Using standard techniques from nuclear physics the averaged matrix element can be expressed as  
\begin{align}
\label{eq:MatrixElement^2}
\langle |\mathcal{M}_{\rm eff}|^2 \rangle &= \frac{4\pi}{2J+1}\sum_{\tau, \tau'} \sum_{L}\left[ R_M^{\tau \tau^{'}} \langle j_i||M_{L;\tau}|j_f\rangle \langle j_f||M_{L;\tau^{'}}||j_i \rangle \right.  \nonumber\\
& \left.+ R_{\Sigma^{''}}^{\tau \tau^{'}}\langle j_i||\Sigma_{L;\tau}^{''}||j_f \rangle \langle j_f ||\Sigma_{L;\tau^{'}}^{''}||j_i \rangle 
  + R_{\Sigma^{'}}^{\tau \tau^{'}}\langle j_i||\Sigma^{'}_{L;\tau}|j_f \rangle \langle j_f||\Sigma^{'}_{L;\tau^{'}}||j_i \rangle 
  \right.  \nonumber\\   
& \left. +\frac{q^2}{m_N^2} R_{\Phi^{''}}^{\tau \tau^{'}}\langle j_i||\Phi^{''}_{L;\tau}||j_f \rangle \langle j_f||\Phi^{''}_{L;\tau^{'}}||j_i \rangle+2 \frac{q^2}{m_N^2}R_{M \Phi^{''}}^{\tau \tau^{'}} \langle j_i||\Phi^{''}_{L;\tau}||j_f \rangle \langle j_f||M_{L;\tau^{'}}||j_i \rangle \right. \nonumber\\
&+\frac{q^2}{m_N^2}R_{\tilde{\Phi}^{'}}^{\tau \tau^{'}} \langle j_i||\tilde{\Phi}^{'}_{L;\tau}||j_f \rangle \langle j_f||\tilde{\Phi}^{'}_{L;\tau^{'}}||j_i \rangle  + \frac{q^2}{m_N^2}R_{\Delta}^{\tau \tau^{'}} \langle j_i||\Delta_{L;\tau}||j_f \rangle \langle j_f||\Delta_{L;\tau^{'}}||j_i \rangle \nonumber \\
& \left.+ \frac{q^2}{m_N^2}R_{\Delta \Sigma^{'}}^{\tau \tau^{'}} \langle j_i||\Sigma^{'}_{L;\tau}||j_f \rangle \langle j_i||\Delta_{L;\tau^{'}}||j_i \rangle \right.  \nonumber\\
&  \left. +\frac{q^2}{m_N^2}R_{\Sigma}^{\tau \tau^{'}} \langle j_i||\Sigma_{L;\tau}||j_f \rangle \langle j_f||\Sigma_{L;\tau^{'}}||j_i \rangle +\frac{q^2}{m_N^2}R_{\Delta^{'}}^{\tau \tau^{'}} \langle j_i||\Delta^{'}_{L;\tau}||j_f \rangle \langle j_f||\Delta^{'}_{L;\tau^{'}}||j_i \rangle \right.\nonumber\\
&\left. +\frac{q^2}{m_N^2}R_{\tilde{\Omega}}^{\tau \tau^{'}}\langle j_i||\tilde{\Omega}_{L;\tau}|j_f \rangle \langle j_f||\tilde{\Omega}_{L;\tau'}|j_i \rangle+\frac{q^2}{m_N^2}R_{\tilde{\Phi}^{'}}^{\tau \tau^{'}}\langle j_i||\tilde{\Phi}_{L;\tau}||j_f \rangle \langle j_f||\tilde{\Phi}_{L;\tau^{'}}||j_i \rangle \right.  \nonumber\\ 
& \left.+ \frac{q^2}{m_N^2} R_{\tilde{\Delta}^{''}}^{\tau \tau^{'}}\langle j_i||\tilde{\Delta}^{''}_{L;\tau}||j_f \rangle \langle j_f||\tilde{\Delta}^{''}_{L;\tau^{'}}||j_i\rangle + \frac{q^2}{m_N^2}R_{\Delta^{'} \Sigma}^{\tau \tau^{'}} \langle j_i||\Delta^{'}_{L;\tau}||j_f \rangle \langle j_f||\Sigma_{L;\tau^{'}}||j_i \rangle
 \right] \;,
\end{align}
where $R_A^{\tau \tau'}$ are DM response functions which are known analytically and are given in Appx.~\ref{Appx:DMresponse}. The nuclear response functions $ W^{\tau \tau'}_{A B} =\sum_L \langle j_i || A_{L,\tau} || j_f \rangle \langle j_f || B_{L,\tau'} || j_j \rangle $ are given by  angular momentum reduced nuclear matrix elements of definite angular momentum $\langle j_f || A || j_i \rangle$. They depend on nuclear matrix elements and have to be evaluated numerically. The operators $A\in \{M, \Sigma', \Sigma'', \Phi'', \tilde{\Phi}', \Delta,\Delta', \Sigma, \tilde{\Omega}, \tilde{\Phi} $ ,$\tilde{\Delta}''\}$  arise from the multipole expansion of the nucleon operators in Tab.~\ref{tab:Operators} and are well established from the treatment of semileptonic weak interactions with nuclei \cite{Donnelly:1978tz,Walecka:1995mi,Serot:1978vj}\footnote{Following the notation of \cite{Fitzpatrick:2012ix} the operator $\tilde{A}$ denotes the symmetrized version of operator $A$.}. The first five lines of eq.~\ref{eq:MatrixElement^2} collect terms which are present in elastic and inelastic DM nucleus scattering and match the squared matrix element given in \cite{Anand:2013yka}. The terms in the last three lines are not considered in the standard non-relativistic theory of DM-nucleus interactions since they vanish for ground state to ground state transitions and are thus irrelevant for elastic scattering. However, they need to be included in an analysis of inelastic scattering.

\begin{figure}[t]
    \centering
        \includegraphics[width=0.49\textwidth]{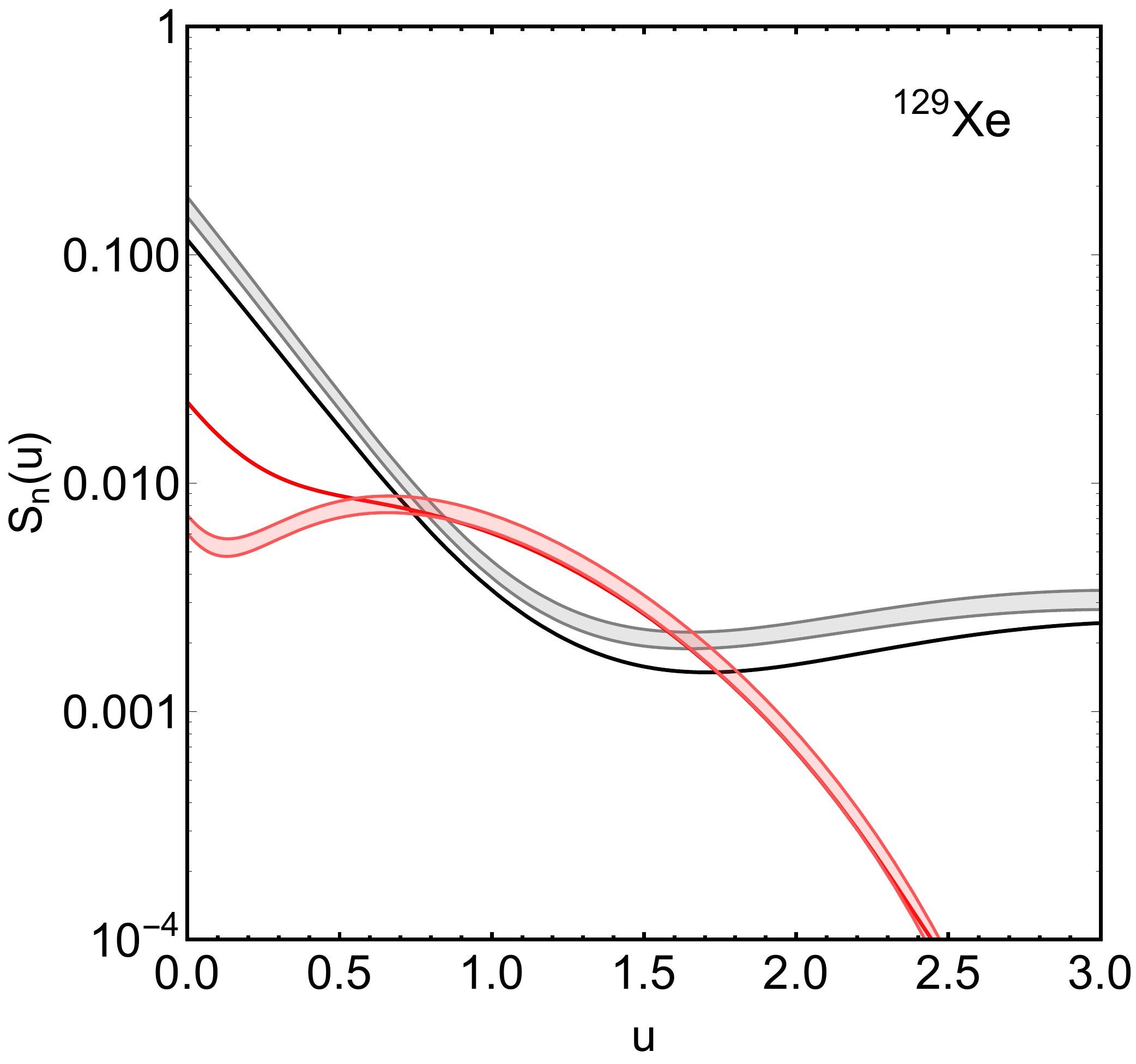} 
        \includegraphics[width=0.49\textwidth]{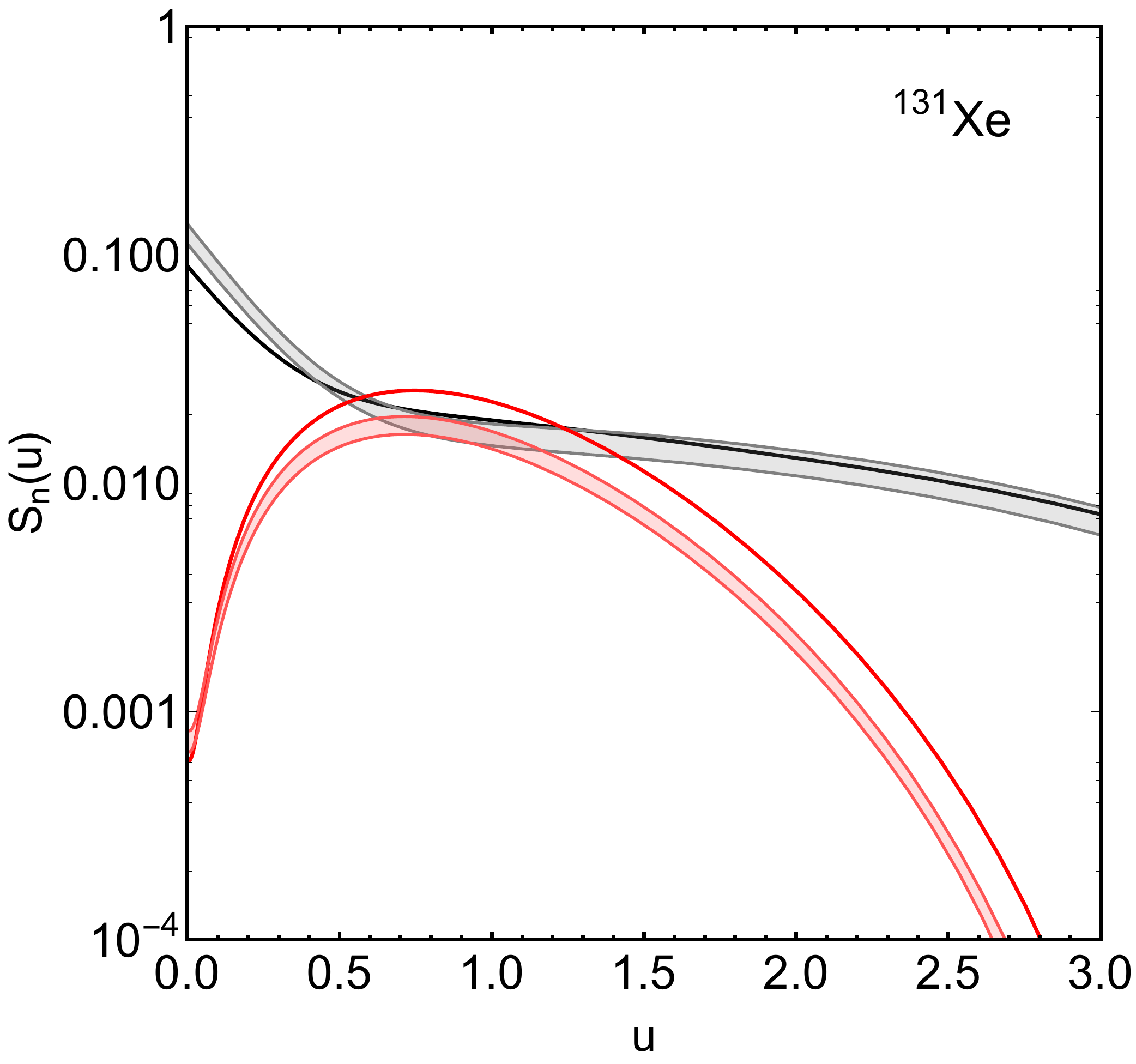}
    \caption{Nuclear structure function for spin-dependent interactions with neutrons $S_n$ as a function of $u= q^2 b^2/2$ for $^{129}$Xe (left) and $^{131}$Xe (right). Our results for (in-)elastic scattering are shown in black (red). For comparison we show the results of \cite{Baudis:2013bba}. The literature result for the inelastic structure function is indicated by the  light red band while the elastic structure function corresponds to the gray band. The width of the bands indicates the estimate of the theoretical uncertainty reported in \cite{Baudis:2013bba}.}
    \label{fig:nuc_comparison}
\end{figure}

Expanding the suppressed indices, the angular momentum reduced matrix elements $\langle j_f || A || j_i \rangle$ can be expressed in terms of the Wigner 3-j symbol and the isospin reduced matrix elements as
\begin{align}
 \langle J_f;T_f M_{T_f} || A_{L,\tau} || J_i;T_i M_{T_i} \rangle=(-1)^{T_f-M_{T_f}} \threej{T_f}{\tau}{T_i}{M_{T_f}}{M_{\tau}}{M_{T_i}} \langle J_f;T_f \vdots\vdots A_{L,\tau} \vdots\vdots J_i;T_i  \rangle\,.
\end{align}
Considering only one-body currents, the nuclear matrix elements can be further simplified. Expressed as the sum of products of single particle matrix elements with the elements of the  one-body density matrix (OBDM) they read 
\begin{align}
    \langle J_f;T_f \vdots\vdots A_{L,\tau} \vdots\vdots J_i;T_i  \rangle =\sum_{|\alpha||\beta|} \psi^{f,i}_{L\tau}(|\alpha|,|\beta|) \langle |\alpha|\vdots\vdots A_{L,\tau}\vdots \vdots |\beta| \rangle
\end{align}
where $|\alpha|$ and $|\beta|$ represent the set of nonmagnetic single-nucleon spatial and spin quantum numbers and the sum extends over the complete set of $|\alpha|$ and $|\beta|$. In the harmonic oscillator basis the  one-body nuclear matrix elements $\langle |\alpha|\vdots\vdots A_{J,T}\vdots \vdots |\beta| \rangle$ of the relevant operators are known analytically, see for example \cite{Catena:2015uha}.  The one-body density matrix elements $\psi^{f,i}_{L;\tau}(|\alpha|,|\beta|)$ parametrize the overlap of the nucleon wave functions in the nucleus. They need to be determined in a nuclear physics computation.
Following the literature 
we perform a nuclear shell model calculation. The contribution from closed shells in the core is given by ~\cite{Walecka:2001gs}
\begin{equation}
\psi_{L;\tau}(|\alpha| ,|\beta|)=\sqrt{2(2J+1)(2T+1)(2 j_\alpha+1)} \delta_{|\alpha||\beta|}\delta_{L0}\,.
\label{eq:Core}
\end{equation} 
Treating the outer shells is more difficult and their contribution can only be determined numerically. We use the public shell model code Nushell@MSU~\cite{Brown:2014bhl}. We perform calculations for $^{129}$Xe and $^{131}$Xe using an $3s_{1/2}2d_{3/2}2d_{5/2}1g_{7/2}1h_{11/2}$ model space above a $^{100}$Sn core, the  ``jj55pn'' model space of Nushell@MSU, and the ``sn100pn'' interaction Hamiltonian~\cite{Brown:2004xk}. For $^{131}$Xe we perform an unrestricted diagonalization of the valence nucleon system. For $^{129}$Xe the basis is too large and we were not able to perform  an unrestricted diagonalization with our computational resources. Therefore, we need to truncate  the model space. Following \cite{Baudis:2013bba} we limit the number of neutron excitations into the energetically disfavored $1h_{11/2}$ orbital to three. A table with the values of OBDMEs in the format suitable for use with the DMFormFactor package is provided as supplementary material.
In order to assess the validity of our shell-model computation we perform a comparison with the literature. We compare our results for the (in-)elastic nuclear structure functions for interactions with neutrons $S_n$ with~\cite{Baudis:2013bba}. As can be seen in Fig.~\ref{fig:nuc_comparison} the overall agreement is good.  Nevertheless,
we would like to caution that our nuclear structure calculation is exploratory and would like to encourage the community to improve it. A more detailed study would be highly beneficial for a future search for inelastic scattering in direct detection experiments.

\section{Inelastic scattering in LXe experiments}
\label{sec:xenonExp}
With the formalism for the description of nuclear excitation induced by DM scattering at hand we move forward by considering the experimental aspects in more detail.
We will first analyze the signal rates associated with the different operators. After a brief introduction into the working principle of LXe TPCs we present the Monte Carlo (MC) simulation that converts the deposited energy of incident particles into measurable detector signals. Finally, we will introduce the relevant background sources and their rates and conclude by presenting the signal-background discrimination in LXe TPCs.

\subsection{Signal rates}
The scattering rates in a detector at earth are given by
\begin{align}
\frac{d R}{d E_R}= \frac{1}{m_T}\frac{\rho_{\chi}}{m_{\chi}}\int_{v\geq v_{min}}d^3v\, v\, f(\vec{v}+\vec{v}_E)\, \frac{d \sigma}{d E_R}
\label{eq:DiffRecoilRate}
\end{align}
where $m_T$ and $m_{\chi}$ are the target and DM mass, respectively. The local DM density is denoted $\rho_{\chi}$  while $\vec{v}$ is the DM velocity. The velocity distribution of DM in the galactic rest frame is given by $f(\vec{v})$ and $\vec{v}_E$ is the velocity of Earth.
We use the Standard Halo model which assumes a truncated Maxwell-Boltzmann distribution for $f(\vec{v})$. The input parameters are the solar circular speed, which we take to be $v_0=220\,\kmps$, the speed of Earth $v_{E}=232 \,\kmps$ and the galactic escape speed $v_{esc}=550\, \kmps$. 

The lower boundary of the integration, $v_{min}$, is the minimal speed required to induce a nuclear recoil with energy $E_R$. It is given by
\begin{align}
    v_{min}=\sqrt{\frac{m_T E_R}{2 \mu^2_T}} + \frac{E^\ast}{\sqrt{2 m_T E_R}}\,,
\end{align}
where $\mu_T$ is the reduced mass of the DM-nucleus system and $E^\ast$ is the nuclear excitation energy.

As mentioned in the introduction, two xenon isotopes, $^{129}$Xe and $^{131}$Xe, have a low lying first excited state.  Both isotopes are very common and have a natural abundance of $26.4\%$ and $21.2\%$ in xenon, respectively. The excitation energy of $^{129}$Xe is $39.6 \,\mbox{keV}$ and the first excited state of  $^{131}$Xe lies at $80.2 \,\mbox{keV}$. While this $\mathcal{O}(10\,\mbox{keV})$ energy difference is small on nuclear scales it is sizable compared to the typical amount of energy available in DM-nucleus scattering and leads to a significant increase of $v_{min}$ compared to elastic scattering. Therefore, we should expect at least a moderate suppression of the inelastic rate with respect to the elastic rate.

It is instructive to take a look at the total rates of inelastic and elastic scattering. We report the ratio of the inelastic rate $R_{in}$ to the elastic rate $R_{el}$ in Tab.~\ref{tab:Ratio} for the representative choice of $m_{\chi}= 300$ GeV and the same coupling to neutrons and protons (isoscalar interaction).
\begin{table}[t]
    \centering
    \begin{tabular}{|c|c|c|c|c|c|c|c|c|c|c|c|c|c|c|c|}
    \hline
        Operator & 1& 3& 4& 5& 6& 7& 8& 9& 10& 11& 12& 13& 14& 15\\
        \hline
         $R_{in}/R_{el}$ & $-$ & $-$ & 0.13 & 0.04 & 0.06 & 62.& 0.009 & 0.07 & 0.11 & $-$& $-$ & 27. & $-$& $-$\\
         \hline
    \end{tabular}
     \caption{Ratio of the inelastic scattering rate on natural xenon $R_{in}$ relative the elastic rate $R_{el}$ for the 14 Operators of the non-relativistic effective theory for $m_{\chi}= 300\,$GeV  and $c^p_k=c^n_k$. For operators marked by "$-$" the ratio is $<10^{-3}$.
     \label{tab:Ratio}
     }
\end{table}
Six of the 14 basic operators predict an inelastic scattering rate  which is less than $10^{-3}$ of the elastic scattering rate. Given that the elastic signal has not been detected yet, the observation of these inelastic signals is challenging and is unlikely to take place in near future. 
 In contrast, the operators, $\mathcal{O}_4$, $\mathcal{O}_5$, $\mathcal{O}_6$,  $\mathcal{O}_9$ and $\mathcal{O}_{10}$ feature an inelastic scattering rate that is only mildly suppressed by $\approx 5 -15 \%$. For those interactions, a future detection of inelastic transitions is conceivable. These very different outcomes for the various operators  
 can be understood from the basic properties of elastic and inelastic scattering. As it is well known, the elastic scattering amplitude of the standard SI interaction, which corresponds to $\mathcal{O}_1$, is  coherently enhanced and scales as $A^2$ where $A$ is the number of nucleons in the nucleus. Consequently, inelastic scattering is strongly suppressed for $\mathcal{O}_1$.
 In contrast, the standard SD interaction, which is described by $\mathcal{O}_4$, does not profit from the $A^2$ enhancement since the interaction leads to a spin-flip. By definition inelastic transitions change the structure of the nucleus and are therefore incoherent
 processes.
 Hence, the inelastic scattering cross section of $\mathcal{O}_4$ is not
 suppressed compared to the elastic one. 
 This pattern can be generalized to the other operators and we find that basically every DM-nucleus interaction that is sensitive to the nuclear spin predicts interesting inelastic transition rates while the other operators do not. For most of these operators $R_{in}/R_{el}=\mathcal{O}(0.1)$ which agrees with the naive reasoning that the higher momentum transfer required by the inelastic process leads to a mild suppression. 
 However, two operators, $\mathcal{O}_7$ and $\mathcal{O}_{13}$, predict that the inelastic rate exceeds the elastic one by one order of magnitude. This effect is more subtle and can ultimately be traced back to the nuclear structure; the elastic scattering rate of $\mathcal{O}_7$ and $\mathcal{O}_{13}$ is suppressed by $v_T^\perp$ while the inelastic rate receives additional contributions from the purely inelastic nuclear response functions  $\tilde{\Omega}$ and $\tilde{\Phi}$ that do not suffer from this suppression.  In the following, we will not consider $\mathcal{O}_8$, which is a borderline case with $R_{in}/R_{el}\approx 10^{-2}$, and focus on the operators with a clear detection potential.
 
 Due to the even proton number in xenon, the dominating contribution to the interaction comes from unpaired neutrons in the two considered isotopes $^{129}$Xe and $^{131}$Xe. In addition, nuclear structure factors for protons suffer from substantial nuclear uncertainties which are known to impact the interpretation of experimental data~\cite{Garny:2012it,Aprile:2013doa}.
In the following, we assume a similar interaction strength of DM with neutrons and protons and will therefore only include the neutron coupling in our analysis.

\subsection{Liquid xenon dual-phase detectors}
\label{sec:Detector}

In this section we assess the experimental observability of the introduced DM-nucleus interaction types.

The most sensitive direct detection DM detectors use the dual-phase time projection chamber (TPC) technology and feature a scalable LXe target that is contained in a cylindrical vessel. Radiation penetrating the target can deposit energy in the form of nuclear recoils (NRs) by scattering off the xenon nucleus (in case of WIMPs or neutrons) or in the form of electronic recoils (ERs) by interacting with atomic electrons (in case of $\gamma$-rays and $\beta$-electrons). The recoils excite and ionize neighbouring xenon atoms resulting in two distinct measurable quantities proportional to the recoil energy. For NRs some energy is lost to heat. The partition into measurable channels depends on the recoil energy and the recoil type. Consequently, the independent measurement of the excitation and ionization signals provides a discrimination between WIMP signals and ER backgrounds. Additionally, the two signals allow for a 3-dimensional reconstruction of the interaction's position which is used to select only events from the inner radio-pure volume of the detector and therefore exploit the excellent self-shielding properties of LXe.

Xenon atoms excited by the recoil form dimers which de-excite on time scales $\mathcal{O}((1-10)\,\mathrm{ns})$ by emitting scintillation light with an average wavelength of $178\,$nm. The light signal (S1 signal) is typically observed by photomultiplier tubes (PMTs) that cover the top and bottom surfaces of the cylindrical volume. The ionization charges are extracted from the interaction site by an electric field ($\mathcal{O}(0.1\,\mathrm{kV/cm})$) applied across the target. The electrons are drifted upwards to the liquid-gas interface where they are extracted into the gaseous xenon phase by a second field ($\mathcal{O}(1\,\mathrm{kV/cm})$). The extraction field accelerates the electrons such that they excite and ionize the gas atoms inducing a second scintillation signal (S2 signal) that is proportional to the number of electrons~\cite{PropScintillation}.

The two lowest energy states of $^{129}$Xe and $^{131}$Xe feature lifetimes $\lesssim 1\,$ns and de-excite by emitting photons with energies of $39.6 \,\mbox{keV}$ and $80.2 \,\mbox{keV}$, respectively. In an inelastic scattering process of WIMPs, the NR and the ER induced by the de-excitation photon cannot be resolved by current detectors due to the short lifetime. Hence, both signals are measured simultaneously resulting in an event characteristic different from the standard elastic NR signal.

\subsection{Detector Simulation}
\label{sec:simulation}

We use a MC simulation in order to convert an energy deposition in LXe into the detectable signals S1 and S2 and follow the same approach as the NEST model \cite{NEST_2011,NEST_2013,NEST_2015}. In the following, we use the symbol $n$ as an average number of quanta and $N$ as the number of quanta sampled from a probability density function. $N$ is therefore impacted by statistical fluctuations. Details on the parameter values used in the simulation can be found in Appx.~\ref{sec:app_simulation}.

The average number of quanta $n_\mathrm{q}$ freed by a recoil of the energy $\epsilon$ divides into the number of excitons $n_\mathrm{ex}$ and the number of ions $n_\mathrm{i}$ and is given by
\begin{equation}
 \begin{aligned}
    n_\mathrm{q} &= L\cdot\frac{\epsilon}{W}\qquad \mathrm{and}\\
    n_\mathrm{q} &= n_\mathrm{ex} + n_\mathrm{i},
    \label{eq:number_quanta}
\end{aligned}
\end{equation}
with the average energy required per quantum $W = (13.7\pm0.2)\,$eV
and the Lindhard quenching factor $L$ that accounts for the energy lost to atomic motion. In order to consider fluctuations in the particle tracks, the number of quanta $N_\mathrm{q}$ is sampled from a Gaussian with the mean $n_{\mathrm{q}}$ and a standard deviation of $\sqrt{F\cdot n_\mathrm{q}}$, where $F$ is the Fano factor \cite{FanoFactor_1995}.
The partition into excitons and ions is determined by
\begin{equation}
 \begin{aligned}
    N_\mathrm{i} &= \mathrm{Binom}\left(N_\mathrm{q},\,\frac{1}{1+n_\mathrm{ex}/n_\mathrm{i}}\right)\qquad \mathrm{and}\\ 
    N_\mathrm{ex} &= N_\mathrm{q}-N_\mathrm{i},
\end{aligned}
\end{equation}
with the exciton-ion-ratio $n_\mathrm{ex}/n_\mathrm{i}$. All created excitons lead to the emission of a scintillation photon. However, ion-electron pairs can recombine, creating additional excitons and at the same time reducing the number of free electrons. The number of recombinations $N_\mathrm{reco}$ depends on the recombination probability $r$ and respective statistical fluctuations:
\begin{equation}
    N_\mathrm{reco} = \mathrm{Gaus}(r N_\mathrm{i},\,\sigma_\mathrm{r} N_\mathrm{i})\,.
\end{equation}
The parameter $r$ is described by the Thomas-Imel box model \cite{ThomasImel} and depends on the applied electric field.
In order to calculate the number of photons $N_\gamma$ emitted from the interaction site, another quenching factor $f_\text{l}$ has to be applied. This factor takes into account that two excitons can interact and produce only a single photon (Penning effects). Consequently, the final measurable quantities $N_\gamma$ and $N_\mathrm{e}$ are given by:
\begin{equation}
 \begin{aligned}
N_\gamma &= \mathrm{Binom}(N_\mathrm{ex},\,f_\mathrm{l})+N_\mathrm{reco} \,, \\
  N_\mathrm{e} &= N_\mathrm{i}-N_\mathrm{reco}\,.
 \end{aligned}
\end{equation}

After modelling the LXe microphysics, the quanta produced at the interaction site are propagated through the detector. The conversion of a scintillation photon into a measurable signal $N^\mathrm{det}_\gamma$ in a PMT is a binomial process
\begin{equation}
    N^\mathrm{det}_\gamma = \mathrm{Binom}(N_\gamma,\,g_1)\,,
\end{equation}
with the detection probability $g_1$ that depends on detector specific parameters such as the geometry, reflectivity of materials, quantum efficiency of the PMTs, etc. On the side of the electron signal, the number of detected electrons $N_\mathrm{e}^\mathrm{det}$ depends on the probability $p_\mathrm{ex}$ of extracting the electrons from the liquid into the gaseous xenon phase:
\begin{equation}
    N_\mathrm{e}^\mathrm{det} = \mathrm{Binom}(N_\mathrm{e},\,p_\mathrm{ex})\,.
\end{equation}
The measured S1 signal in number of photo-electrons (PE) emitted from the PMT photo-cathode is given after taking into account the resolution $\sigma_\mathrm{PMT}$ of the PMT response to photons:
\begin{equation}
    \mathrm{S1} = \mathrm{Gaus}(N_\gamma^\mathrm{det},\,\sigma_\mathrm{PMT}\sqrt{N_\gamma^{det}})\,.
\end{equation}
For S2 signals, the PMT response is not limiting the resolution due to the larger number of photons ($\mathcal{O}(10)$) generated during the proportional scintillation process. However, this process is also a statistical one and the electron amplification gain $\omega$ with its standard deviation $\sigma_\omega$ is determining the measured S2 signal:
\begin{equation}
    \mathrm{S2} = \mathrm{Gaus}(\omega N_\mathrm{e},\,\sigma_\omega\sqrt{N_\mathrm{e}}).
\end{equation}

For NRs the procedure is followed exactly in the described way and the parameters are set to the values given in Tab.~\ref{tab:LxeParams}. For ERs the expression for the recombination probability is somewhat more complicated and depends on both, the electric field and the deposited energy $\epsilon$. Furthermore, incident gamma and beta particles show small differences in the partition of the deposited energy into the mean number of photons and electrons emitted and extracted from the interaction site, $n_\gamma$ and $n_\text{e}$, respectively. These differences are caused by the fact that beta particles only induce recoils on the outer shell electrons while gamma particles can also interact with inner shell electrons. The latter process results in additional Auger electrons. NEST implements the complete process and predicts $n_\gamma$ and $n_\mathrm{e}$ in dependence of $\epsilon$ and for various electric field strengths in \cite{NEST_2013} (Figure 1). For simplicity the provided data is utilized in this work and is smeared by the statistical fluctuations of the LXe microphysics processes. The total number of quanta produced at the interactions site is given by $n_\mathrm{q} = n_\mathrm{e} + n_\gamma$ which is valid for ERs due to the non-existing quenching ($L=1$). Furthermore,  $n_\gamma$ and $n_\mathrm{e}$ enter into the calculation of $r$ (see Tab.~\ref{tab:LxeParams}). Given $n_\mathrm{q}$ and $r$ the full simulation including all statistical processes can be run for ERs as described above.

During inelastic scattering of WIMPs, a NR is induced together with an ER from the de-excitation photon. The signals from both recoils cannot be resolved in time. Hence, the S1 and S2 signals from the two processes are added at the end of this simulation resulting in S2/S1-ratios characteristic for inelastic scattering (see Fig.~\ref{fig:Discrimination}).

During the final phase of this work, the new version NESTv2 became available with updated charge and light yield curves for ERs~\cite{NESTv2}. The predicted new curves agree with the previous version within 20\%. Due to time constraints we have not been able to implement the NEST update but we do not expect a large impact on our conclusions.

\subsection{Detector configuration}

Beside the reliance on the knowledge on LXe microphysics, the above described detector simulation depends on the detector configuration given by the electric field $E$, photon detection probability $g_1$, electron extraction probability $p_\mathrm{extract}$, electron amplification gain $\omega$ with its width $\sigma_\omega$, and the PMT single photon resolution $\sigma_\mathrm{PMT}$.

The most sensitive xenon detectors constructed to date are the XENON1T~\cite{Xe1T-SI}, LUX~\cite{LUX-SI} and PandaX-II~\cite{PandaX-SI} experiments. An overview on the respective values for the mentioned parameters is given in Tab.~\ref{tab:DetectorParameters}. While the anticipated electric drift field across the TPC for XENON1T and LUX was $1$ and $2\,$kV/cm, respectively \cite{xe1t_physicsreach,LUX-Exp}, the achieved values were lower. However, higher drift fields have been used by the PandaX-II ($400\,$V/cm) and previous experiments such as XENON100 ($530\,$V/cm)~\cite{xe100_SI} and ZEPLIN-III ($3400\,$V/cm)~\cite{zeplin}. The NEST model suggests that the discrimination between NRs and ERs is improved for larger electric fields, especially at high energies \cite{NESTv2}. Hence, future detectors are still aiming at high electric drift fields of several hundred V/cm \cite{lz_proj,xe1t_physicsreach}. We choose for our benchmark detector model a drift field of $500\,$V/cm. All other parameters are fixed to the values achieved in XENON1T since this experiment has shown the highest sensitivity to WIMP-nucleon interactions among LXe direct detection experiments to date~\cite{Xe1T-SI,xe1t_SD}.

\begin{table}
    \centering
\begin{tabular}{c|c|c|c}
    Parameter & XENON1T & LUX & PandaX-II \\\hline 
    Target mass [kg] & 2000 \cite{Xe1T-SI} & 250 \cite{LUX-SI} & 580 \cite{PandaX-SI}\\
    Exposure [t$\cdot$yr] & 1 \cite{Xe1T-SI} & 0.09 \cite{LUX-SI} & 0.15 \cite{PandaX-SI}\\
    $E$ [V/cm] & 81 \cite{Xe1T-SI} & 181 \cite{LUX-SI} & 400\cite{PandaX-SI} \\
  $g_1$  [PE/$\gamma$]  & $\sim0.14$ \cite{xe1t_analysis} & $\sim0.12$ \cite{LUX_analysis}  & $\sim0.11$ \cite{PandaX-SI}\\
  $\omega$ [PE/e$^-$] & $\sim28$ \cite{xe1t_analysis} & $\sim25$ \cite{LUX_analysis} & $\sim24$ \cite{PandaX-SI} \\
  $\sigma_\omega$ [PE/e$^-$] & $\sim7$ \cite{xe1t_analysis} & $\sim6$ \cite{LUX_analysis} & $\sim7$ \cite{PandaX_Commissioning} \\
  $p_\mathrm{extract}$ [\%] & $\sim93$ \cite{xe1t_analysis} & $\sim50$ \cite{LUX_analysis}& $\sim55$ \cite{PandaX-SI}\\
  $\sigma_\mathrm{PMT}$ [PE/$\gamma$] & $\sim0.4$ \cite{xe1t_PMTs} & $\sim0.3$ \cite{lux_PMTs} & $\sim0.4$ \cite{xe1t_PMTs}\\
\end{tabular}
    \caption{Overview on detector specific parameters in the most sensitive dual-phase LXe detectors to date.}
    \label{tab:DetectorParameters}
\end{table}

It is common within the community to use only the S2 signal S2$_\mathrm{b}$ measured in the bottom PMT array for the WIMP analysis, instead of the full S2 signal. This is due to the production of the S2 signal in the xenon gas phase located a few centimeters below the top PMT array. While the signal is only detected by a few photo sensors in the top array, it distributes more uniformly on the bottom array. Hence, S2$_\mathrm{b}$ is less affected by localized effects from non-functional PMTs. We follow this approach in our simulation and multiply $\omega$ and its width $\sigma_\omega$ by a factor of 0.4 which is the fraction of light observed in the bottom PMT array in XENON1T \cite{xe1t_analysis}.

We simulated the detector threshold by removing all simulated events with $\mathrm{S1}<3\,$PE and $\mathrm{S2}<200\,$PE which is compatible with the region of interest for DM searches with recent LXe experiments \cite{Xe1T-SI,LUX-SI,PandaX-SI}. Even though the signal detection efficiency also highly depends on the performance of the data selection, we do not consider any cut acceptances in the simulation. The combined cut acceptance usually has no strong energy dependence and takes values of around $90\,$\%. Therefore, it could be also treated as a universal scale factor on the signal rate which has small impact on the findings in this work.

\subsection{Backgrounds}
\label{sec:Backgrounds}

The background in large scale LXe direct detection experiments is composed of ERs and NRs induced by ambient radioactivity and solar neutrinos.
The following background components are the most relevant and challenging ones for present and upcoming DM searches:
\begin{itemize}
    \item \textbf{Materials:} $\gamma$ and $\beta$ radiation from the detector materials induce ERs in the target. Tonne scale direct detection experiments are able to exploit the good self-shielding properties of the LXe by selecting an inner radio-pure fiducial volume, reducing this background to a negligible level.
    \item \textbf{$^{85}$Kr:} A small remnant of natural Krypton with the $\beta$-emitting isotope $^{85}$Kr remains in xenon that has been commercially produced by air liquefaction. However, its concentration can be further reduced to a sub-dominant level by cryogenic distillation~\cite{xe1t_distilllation}.
    \item \textbf{$^{222}$Rn:} The radio-active noble gas $^{222}$Rn is produced within the uranium decay chain, emanates from the detector materials and mixes within the LXe target due to its long half-life of 3.8 days. The daughter isotope $^{214}$Pb decays via the emission of a $\beta$-electron contributing to the ER background in the low-energy region. During detector construction, materials are screened and selected for their low intrinsic radioactivity and Radon emanation~\cite{xe1t_screening,pandaX_screening}, and $^{222}$Rn can be removed from LXe by means of cryogenic distillation \cite{xe100_rnremoval,hexe_distillation} or adsorption on charcoal~\cite{lz_adsorption}.
    \item \textbf{$^{136}$Xe:} The isotope $^{136}$Xe is a double $\beta$ emitter with a half-life of $2.17\cdot10^{21}\,$y~\cite{exo_xe136} and a $Q$-value of $2458\,$keV. Its abundance in natural xenon yields 8.9\% and its depletion offers one possibility to reduce this background.
    \item \textbf{Solar $\nu$s:} Solar neutrinos can scatter elastically off electrons, inducing ERs and presenting an irreducible background which is homogeneously distributed over the target.
    \item \textbf{Neutrons:} Neutrons are produced by spontaneous fission and ($\alpha$,n)-reactions within the Uranium and Thoron decay chains. The particles are emitted from the detector materials and propagate through the target. In large scale detectors the vast majority of neutrons ($\sim80$\%) scatter twice within the volume~\cite{xe1t_analysis}. Hence, this background can be reduced by single scatter event selection. Furthermore, future detectors will employ neutron vetos in order to further reduce this background component~\cite{LZ}. Muon-induced neutrons are reduced to a negligible level in large scale detectors by active Cherenkov muon vetos.
    \item \textbf{CE$\nu$NS:} Coherent scattering of $pp$ and $^7$Be solar neutrinos off nuclei induces NRs in the region of interest and represents an irreducible background which ultimately limits the sensitivity of future LXe DM detectors~\cite{Billard_neutrinofloor}.
\end{itemize}

The energy spectra of the discussed background components and respective uncertainties are extracted from the XENON1T prediction~\cite{xe1t_physicsreach}. An exception is the uncertainty of the $^{136}$Xe background rate where we neglect shape uncertainties of the spectrum potentially induced by the phase space factor that were considered in the reference. Instead, we assume that the dominating uncertainty is coming from the half-life measurement which yields 3\%.
The relative rate uncertainties for materials, $^{85}$Kr, $^{222}$Rn, solar neutrinos, neutrons and CE$\nu$NS yield 10\%, 20\%, 10\%, 2\%, 17\% and 17\%, respectively.

\begin{figure}
    \centering
    \includegraphics[width = 0.6\textwidth]{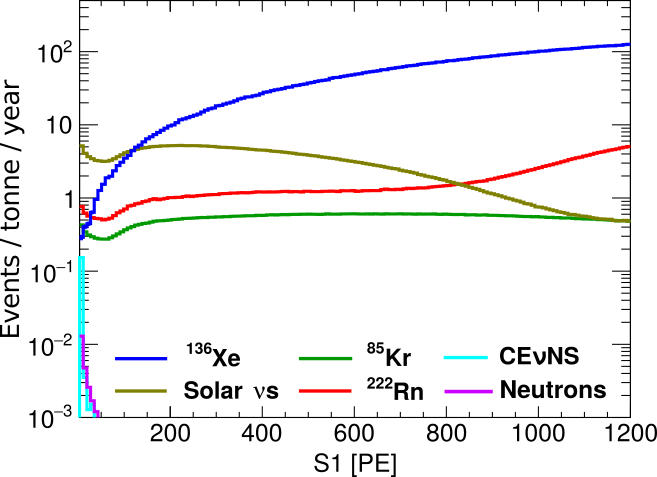}
    \caption{Simulated S1 spectrum for background components in a XENONnT-like detector. Backgrounds from detector artifacts are neglected.}
    \label{fig:Backgrounds}
\end{figure}

In this article we derive prospects for future LXe DM detectors like XENONnT, LZ~\cite{LZ} and DARWIN~\cite{DARWIN} which will have reduced background levels achieved by exploiting the reduction methods explained above. In particular, we consider for a XENONnT-like experiment a reduction of the $^{222}$Rn induced background by a factor of 100 and for the $^{85}$Kr component a factor of 10 w.r.t. the XENON1T prediction, as done in~\cite{xe1t_physicsreach}. Additionally, the background from neutrons is scaled down by a factor of 20 assuming a good performance of the neutron veto and the double scatter rejection. Fig.~\ref{fig:Backgrounds} shows the simulated S1 spectrum for all background components. In the standard WIMP search region (up to ~100 PE) solar neutrinos will be the dominating background while the leading contribution is coming from $^{136}$Xe at higher energies. For a DARWIN-like detector, radiogenic neutrons are assumed to be negligible while the $^{222}$Rn and $^{85}$Kr contributions are conservatively set to the XENONnT level. 

Backgrounds from detector artifacts such as events from accidentally paired lone S1 and lone S2 signals from light and charge insensitive regions or radio-activity at the TPC surfaces that suffer from charge loss and are mis-reconstructed inside the FV are neglected in this study.



\subsection{Signal and background discrimination}

The advantage of LXe TPCs is their ability to measure both the light and charge signal of the interaction. This permits a three dimensional position reconstruction which is a  powerful tool for background reduction by fiducialization as explained above. Furthermore, the ratio between S2 and S1 signals of NRs is smaller than for ERs which allows to differentiate between the two processes and therefore provides a further background discrimination.

\begin{figure}
    \centering
    \includegraphics[width = 0.48\textwidth]{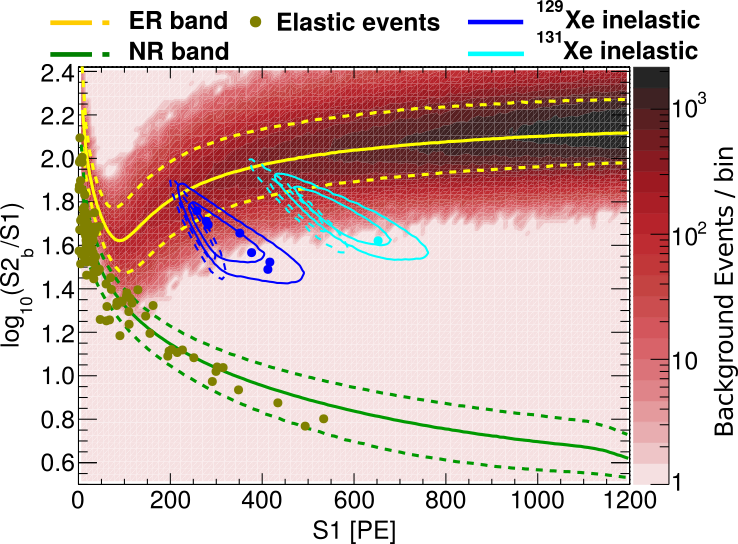}\hspace{0.02\textwidth}
    \includegraphics[width = 0.48\textwidth]{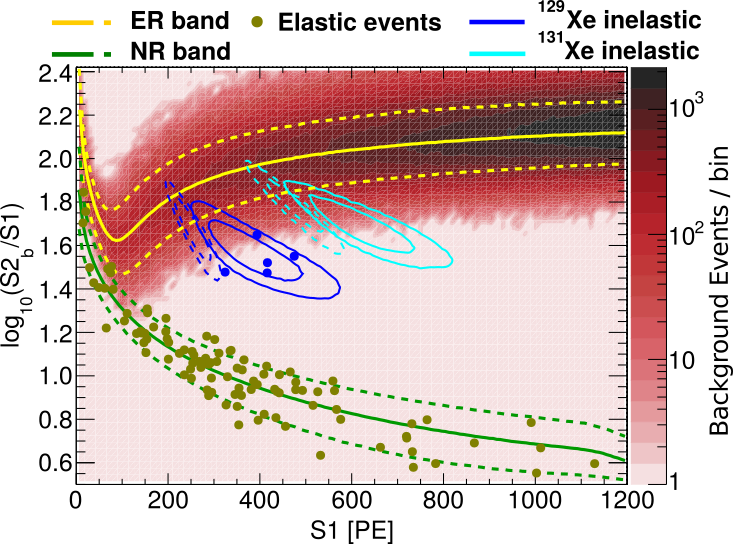}
    \caption{Background and signal distribution in analysis space for a 100\,tonne$\times$year exposure in a XENONnT-like detector and for a $\mathcal{O}_{4}$ (left) and $\mathcal{O}_{6}$ (right) signal with an operator scale (cross-section) of $\Lambda = 1900\,\mbox{GeV}$ ($\sigma_{SD} = 1.5 \times 10^{-41} \,\mbox{cm}^2$) and $\Lambda = 230 \,\mbox{GeV}$, respectively. We assume $m_\mathrm{\chi} = 300\,$GeV. The density of the background distribution is indicated by the color scale. The 10\%-50\%-90\% (dashed-solid-dashed) contours of the ER (NR) regions are indicated by yellow (green) lines. Elastic WIMP-nucleon scattering events are indicated by olive markers and populate the NR band. Solid blue (cyan) lines show the 60\% and 90\% contours of the signal region for WIMPs scattering inelastically off $^{129}$Xe ($^{131}$Xe). Respective markers indicate inelastic DM signal events for the given exposure and cross section. Dashed blue (cyan) lines show the 60\% and 90\% contours of the pure $^{129}$Xe ($^{131}$Xe) de-excitation signal without NR.}
    \label{fig:Discrimination}
\end{figure}

Fig. \ref{fig:Discrimination} shows the background and signal distribution for the effective operators $\mathcal{O}_{4}$ (left) and $\mathcal{O}_{6}$ (right) in log$_{10}$(S2$_\mathrm{b}$/S1) versus S1, denoted as analysis space. We assume a 100\,tonne$\times$year exposure and an effective scale of $\Lambda = 1900\,\mbox{GeV}$ (cross section of $\sigma_{SD} = 1.5 \times 10^{-41} \,\mbox{cm}^2$) and $\Lambda = 230 \,\mbox{GeV}$. The WIMP mass is set to $m_\mathrm{\chi} = 300\,$GeV. This corresponds to a signal of $\simeq 100$ events in the detector. The density of the background population is indicated by the color scale. Regions where NR and ER events are expected to be located are indicated by 10\%-50\%-90\% (dashed-solid-dashed) contours lines and form band-shaped structures. Elastic WIMP signal events (olive markers) populate the NR band that is separated from the ER background.
The mono-energetic de-excitation signals of $^{129}$Xe and $^{131}$Xe (dashed blue and cyan ellipses) are shifted towards smaller S2$_\mathrm{b}$ values compared to the ER band that is populated by $\beta$-radiation induced events. This is due to different photon and electron yields for $\beta$ and $\gamma$ interactions in LXe~\cite{NEST_2015}. For WIMPs scattering inelastically off the two xenon isotopes, the de-excitation signal is added to the NR, resulting in signal regions that are stretched towards higher S1 signals (solid blue and cyan lines). The expected number of inelastic signal events is indicated by blue and cyan markers. The displacement of the signal regions from the ER band allows to suppress the bulk of the background population and therefore enhances the sensitivity. It is interesting to note that the irreducible background for inelastic DM-nucleus scattering, that is inelastic neutrino-nucleus scattering,  is highly suppressed \cite{Pirinen:2018gsd}. 
Therefore, a detector that allows for a clear separation between electronic backgrounds and the inelastic DM signal, which consists of a coincident photon and nuclear recoil, could in principle operate in background free mode. 
Hence, sensitivities  beyond the conventional neutrino floor~\cite{Billard_neutrinofloor} could be achieved.

The recoil spectrum for the effective operator $\mathcal{O}_{4}$ falls off exponentially towards higher energies while the spectrum for $\mathcal{O}_{6}$ features a maximum at $\sim$130\,keV. Hence, elastic events for $\mathcal{O}_{4}$ populate only the low-S1 region while the signal population extends more towards higher S1 values for $\mathcal{O}_{6}$ making this signal well distinguishable from $\mathcal{O}_{4}$ and background also in the purely elastic scattering channel.

\section{Discovery reach for inelastic scattering}
\label{sec:DiscoveryReach}
In this section we study the observability of the inelastic signal contributions for the effective operators $\mathcal{O}_{\alpha}$, $\alpha\in\{4,5,6,7,9,10,13\}$, and a given detector configuration. We use frequentist statistics to derive the DM scattering rate that allows to discriminate the signal from backgrounds with a significance of at least $3\sigma$ in $90\%$ of experiments.

\subsection{Statistical treatment}
 We employ a binned Log-Likelihood procedure for the detection of a positive signal as described in \cite{Cowan:2010js} and \cite{Agashe:2014kda}.
Using the results of Sec.~\ref{sec:xenonExp} we derived the binned probability distributions $f$ for the DM signal, which depend on $m_\chi$ and the operator $\mathcal{O}_\alpha$, and the backgrounds~$\beta$ 
\begin{align}
 f_{\mathcal{O}_{\alpha},m_{\chi}}&:S1-\log_{10}(S2/S1)\to[0,1],\\
 f_{\textrm{R}_{\beta}}&: S1-\log_{10}(S2/S1)\to [0,1],
\end{align}
 with $\beta\in\{\textrm{Solar }\nu\textrm{s}, \,^{85}\textrm{Kr},\, ^{136}\textrm{Xe},\,^{222}\textrm{Rn},\,\textrm{Neutrons},\,\textrm{CE}\nu\textrm{NS}\}$. We simulate $10^6$ events to determine each probability distribution and employ $100\times150$ bins in $S1-\log_{10}(S2/S1)$ space. The theoretically expected number of events $\mu_i$ in the $i$th bin, $i=1,...,N_{\textrm{bins}}$, is given by
\begin{align}
 \mu_i(R_{\chi},\textbf{R}_{\textrm{BG}})=\textrm{Exposure}\times\left(R_{\chi}\cdot f_{\mathcal{O}_{\alpha},m_{\chi}}(i)+\sum_\beta R_{\beta}\cdot f_{\textrm{R}_{\beta}}(i)\right).
 \label{eq:EV}
\end{align}
Here $f(i)$ is the probability of finding an event in the $i$th bin and $R_{\beta}$ and $R_{\chi}$ are the background and the DM rates, respectively.
We denote the different background rates collectively as $\textbf{R}_{\textrm{BG}}:=(R_{\textrm{Solar }\nu\textrm{s}},R_{^{85}\textrm{Kr}}, R_{^{136}\textrm{Xe}},R_{^{222}\textrm{Rn}},R_{\textrm{Neutrons}},R_{\textrm{CE}\nu\textrm{NS}})$ in the following.

For a given set of data $\mathcal{D}$ the binned likelihood function is given by
\begin{align}
 \mathscr{L}(\mathcal{D}|R_{\chi},\boldsymbol{R}_{\textrm{BG}}):=
 \prod_{i=1}^{N_{\textrm{bins}}}
  \left[\frac{(\mu_i(R_{\chi},\textbf{R}_{\textrm{BG}}))^{n_i(\mathcal{D})}}{n_i(\mathcal{D})!}\right]e^{-\mu_i(R_{\chi},\textbf{R}_{\textrm{BG}})}\cdot
  \prod_{\beta}\mathscr{L}_{\beta}(R_{\beta}),
 \label{eq:LF}
\end{align}
where $n_i(\mathcal{D})$ is the observed number of  events in bin $i$. The background  likelihood function
\begin{align}
 \mathscr{L}_{\beta}(R_{\beta}):=\frac{1}{\sqrt{2\pi\sigma_{\beta}^2}}\cdot \exp\left\{{\frac{\left(R_{\beta}-\overline{R}_{\beta}\right)^2}{2\,\sigma_{\beta}^2}}\right\}
\end{align}
parametrizes our knowledge about the expected background rates $\overline{R}_{\beta}$ and allows for deviations $\sigma_{\beta}$  according to the background uncertainties (see Sec.~\ref{sec:Backgrounds}). The likelihood function is maximized both under the null hypothesis of no DM signal and by leaving all rates unconstrained.  The ratio of the maxima
\begin{align}
 \lambda(0):=\frac{\max\limits_{\textbf{R}_{\textrm{BG}}}\mathscr{L}(\mathcal{D}|R_{\chi}=0,\textbf{R}_{\textrm{BG}})}{\max\limits_{R_{\chi},\textbf{R}_{\textrm{BG}}}\mathscr{L}(\mathcal{D}|R_{\chi},\textbf{R}_{\textrm{BG}})}
\end{align}
defines the frequentist test statistic for discovery of a positive signal
\begin{align}
 q_0:=\begin{cases}
       -2\ln\lambda(0) & R_{\chi} \ge 0\\
       0 &R_{\chi} <0
      \end{cases}.
\end{align}
According to Wilks' theorem \cite{Wilks:1938dza} the test statistic approaches a chi-square distribution for an infinite amount of experimental trials if the null hypothesis is true. Combined with the results found by Wald \cite{10.2307/1990256} this allows to reject the background only hypothesis with a significance of $Z=\sqrt{q_0}$ for data $\mathcal{D}$ including a DM signal.

We define the discovery reach for each DM mass as the value of the DM rate $\overline{R}_{\chi}$ for which  $90\%$ of experiments find a $q_0$-value with a statistical significance of $Z\ge3$.  For fixed exposure and DM mass we conduct $2500$ pseudo-experiments and generate mock data  
which we use to  derive the $q_0$ distribution. 
We vary the rate to determine the rate $\overline{R}_{\chi}$ for which $Z\ge3$ for  $90\%$ of the $2500$ $q_0$-values in the set $\{q_0\}$. By repeating this procedure for different DM masses we are able to map out the discovery reach.
In order to ensure the comparability of the elastic and inelastic case we derive both discovery reaches ourselves and  do not rely on results from literature.

\subsection{Results}

\begin{figure}[t]
    \centering
    \includegraphics[width=7.4cm, height=6.95cm]{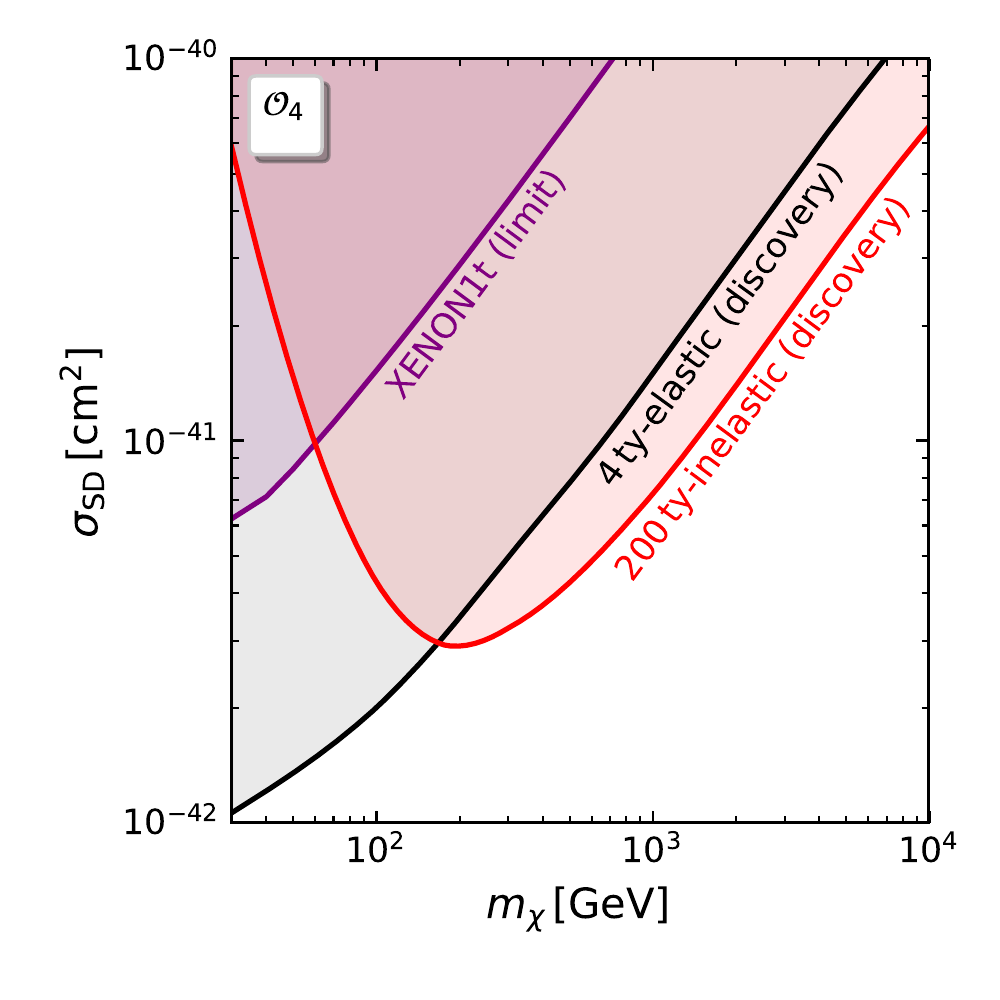} \qquad \includegraphics[width=0.45\textwidth]{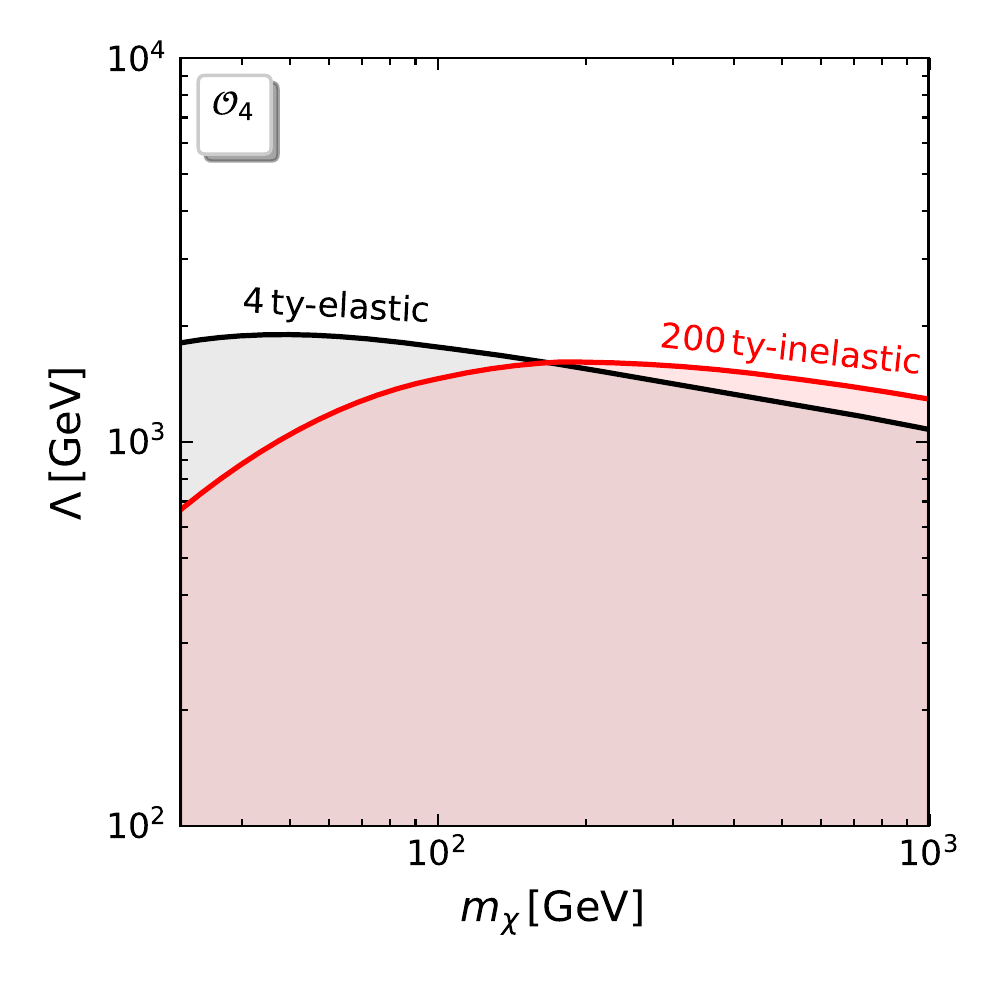}
    \caption{Discovery reach for $\mathcal{O}_{4}$ (standard SD). The left panel shows the reach in the customary $m_{\chi}$-cross section plain and the right panel in terms of $\Lambda$. The black line shows the discovery reach of a search for elastic scattering with a XENONnT-like detector and an exposure of $4 \,\mbox{tonne}\times \mbox{years}$ while the red line is the reach for inelastic scattering with a DARWIN-like detector with a $200 \,\mbox{tonne}\times \mbox{years}$ exposure. The current 90\% CL exclusion limit from XENON1T is indicated by the purple line \cite{xe1t_SD}. }
    \label{fig:DiscoveryReachSD}
\end{figure}

\begin{figure}[t]
    \centering
        \includegraphics[width=0.45\textwidth]{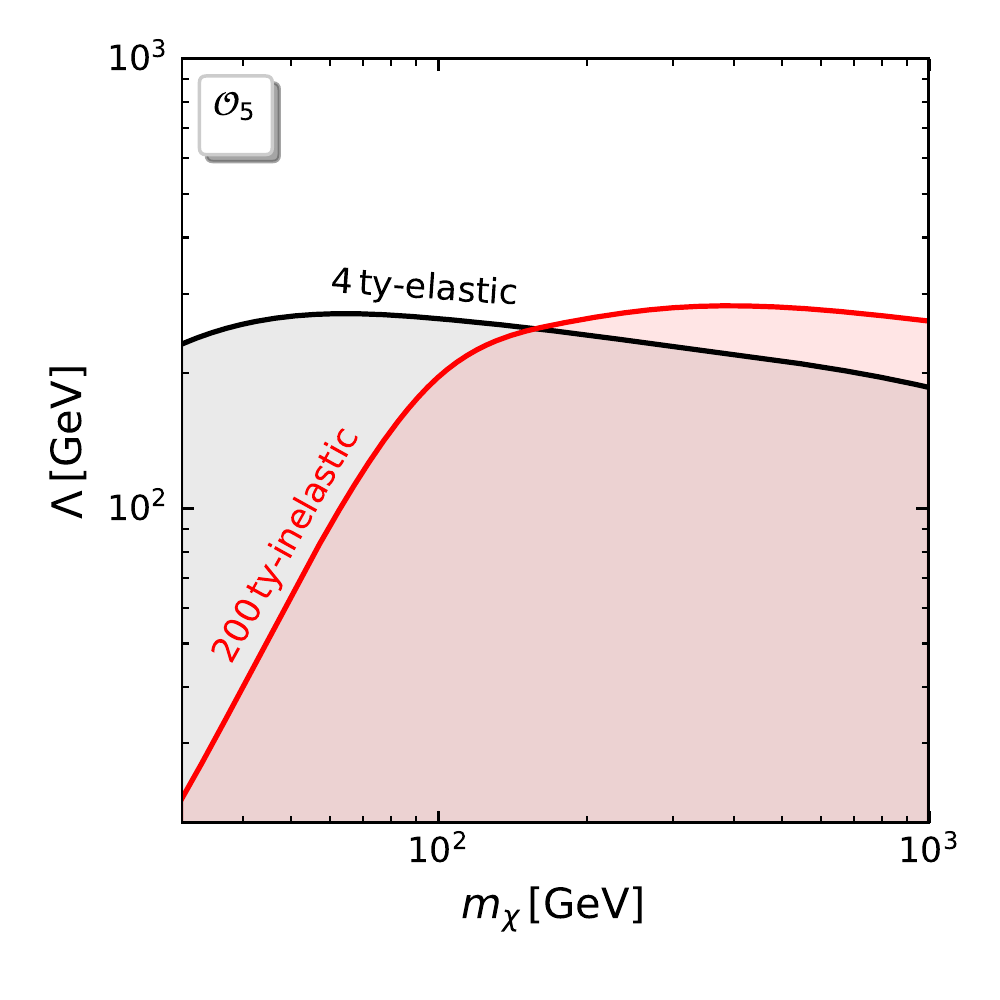} \qquad \includegraphics[width=0.45\textwidth]{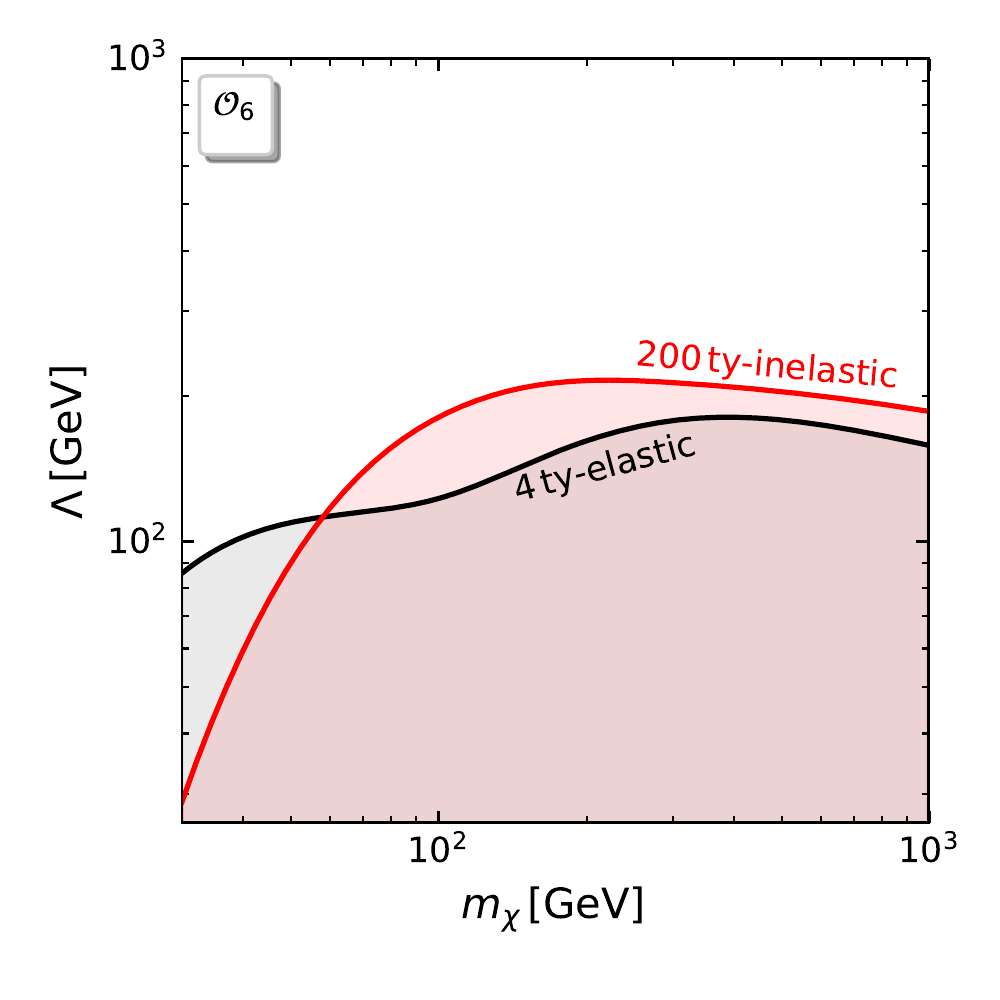}
        \includegraphics[width=0.45\textwidth]{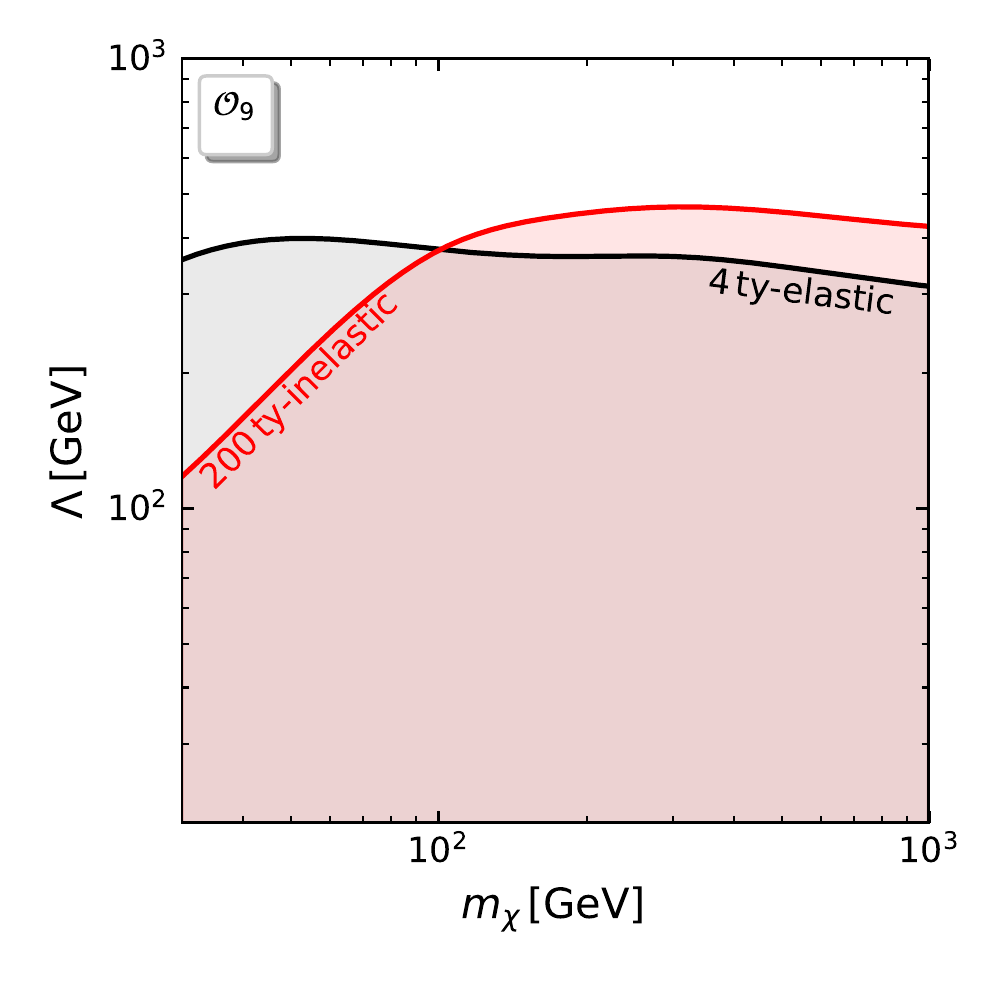}\qquad 
        \includegraphics[width=0.45\textwidth]{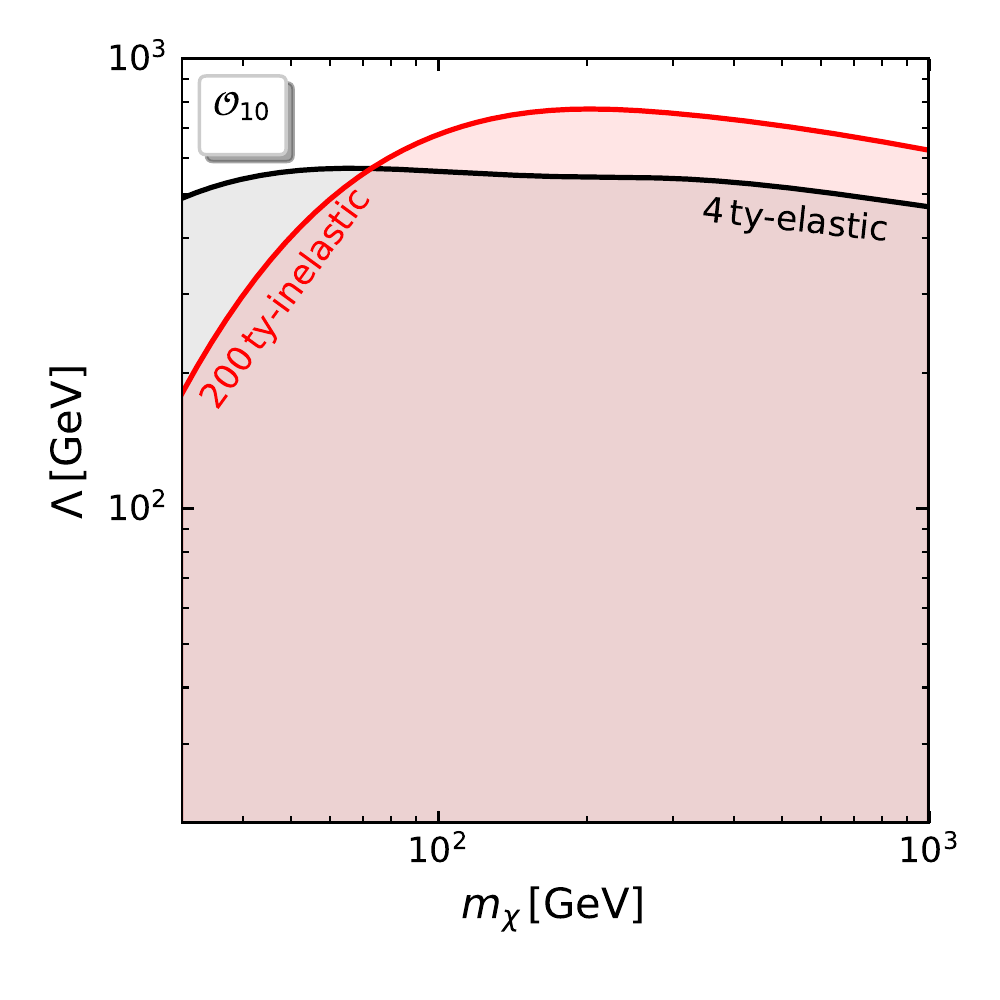}
    \caption{Same as figure~\ref{fig:DiscoveryReachSD} (right panel) for $\mathcal{O}_{5}$, $\mathcal{O}_{6}$, $\mathcal{O}_{9}$ and $\mathcal{O}_{10}$. }
    \label{fig:DiscoveryReachOp}
\end{figure}

\begin{figure}[t]
    \centering
        \includegraphics[width=0.45\textwidth]{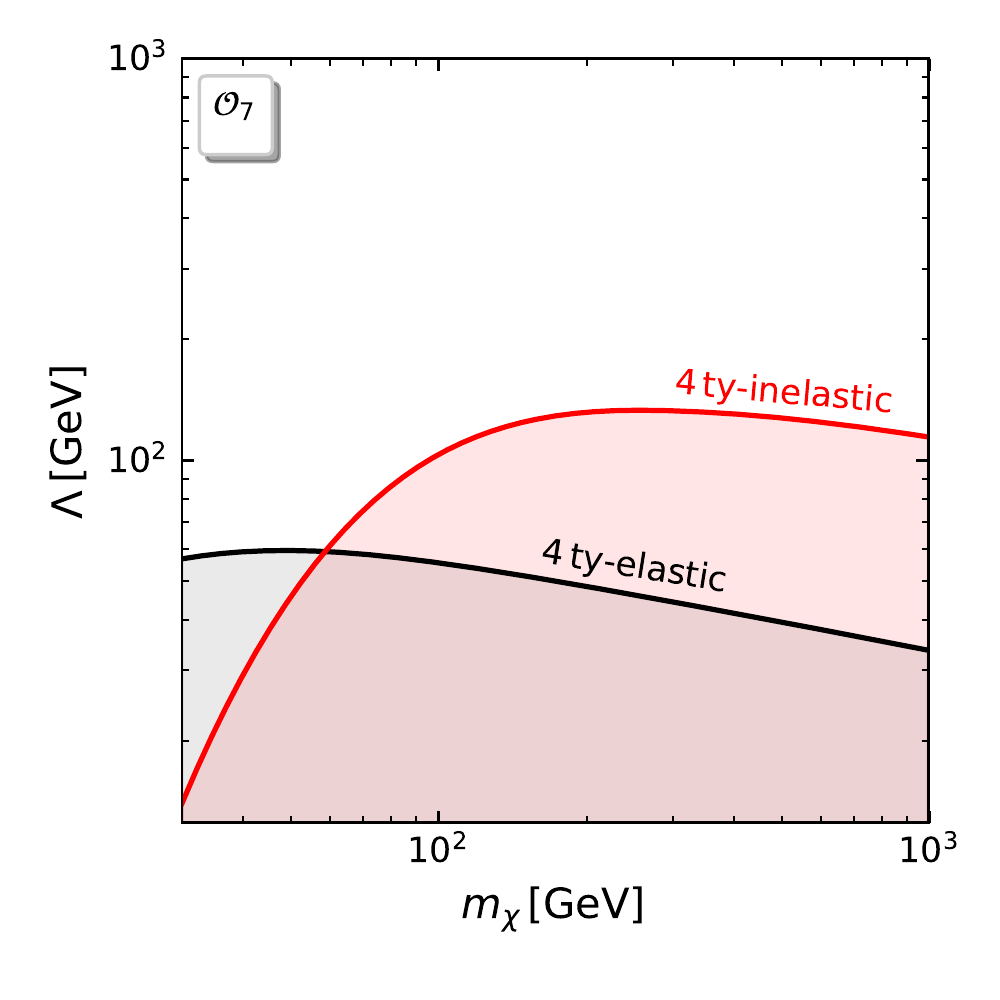} \qquad \includegraphics[width=0.45\textwidth]{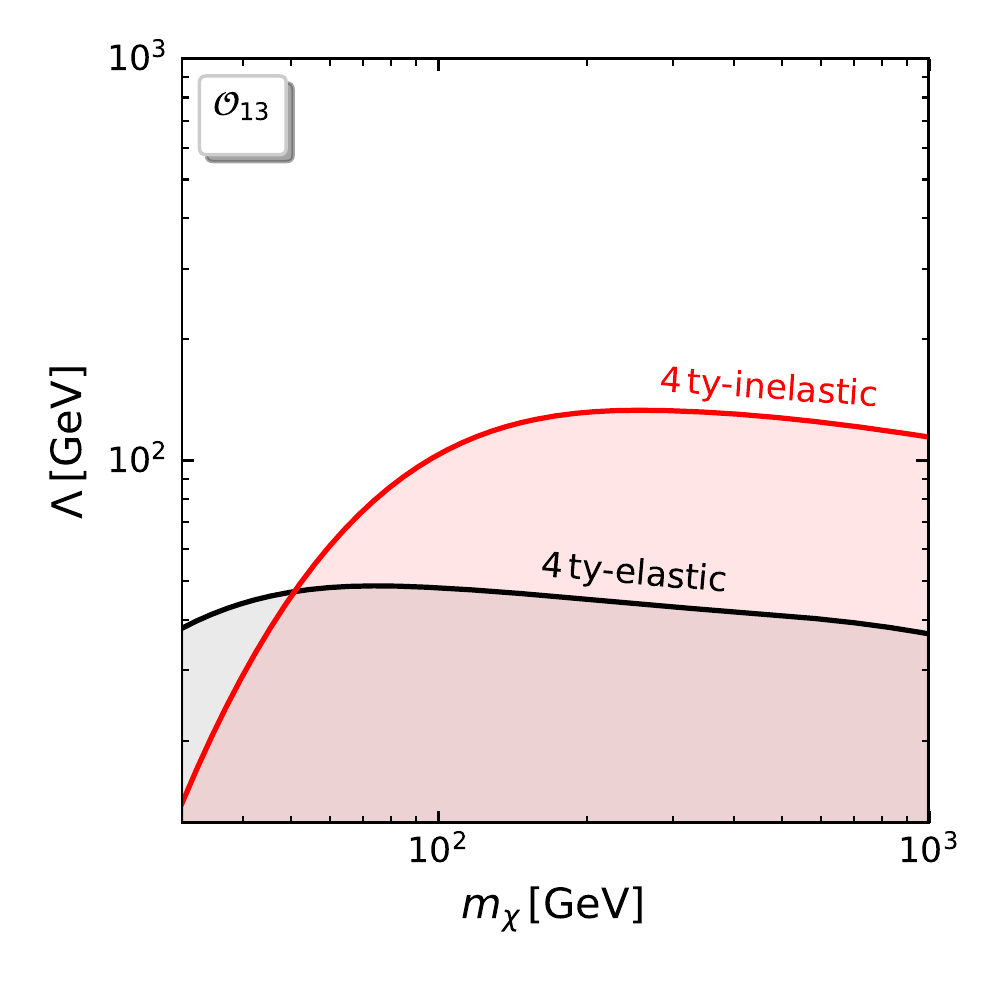}
    \caption{Discovery reach for $\Lambda$ as a function of the DM mass for $\mathcal{O}_{7}$ (left) and $\mathcal{O}_{13}$ (right). The (in-)elastic channel is indicated in black (red) for a XENONnT-like detector with $4$ tonne$\times$years of exposure.}
    \label{fig:DiscoveryReachOp7p13}
\end{figure}

We summarize the main results of this section in Fig.~\ref{fig:DiscoveryReachSD}, Fig.~\ref{fig:DiscoveryReachOp} and Fig.~\ref{fig:DiscoveryReachOp7p13}. We start by reviewing the discovery potential of inelastic transitions induced by operator $\mathcal{O}_4$, which corresponds to DM with the usual spin-dependent interaction, before turning to $\mathcal{O}_5$, $\mathcal{O}_6$, $\mathcal{O}_9$ and $\mathcal{O}_{10}$, all of which have a substantial but subdominant inelastic scattering rate. Finally, we discuss $\mathcal{O}_7$ and $\mathcal{O}_{13}$ which lead to a signal that is dominated by inelastic scattering.

The left panel of  Fig.~\ref{fig:DiscoveryReachSD} displays the discovery reach for elastic (inelastic) scattering of DM on natural xenon assuming the standard SD operator $\mathcal{O}_4$ for future direct detection experiments, i.e. a XENONnT-like (DARWIN-like) detector with 4 tonne$\times$years ($200$ tonne$\times$years) exposure. The current limit on this interaction type from XENON1T ~\cite{Aprile:2019dbj} is depicted in purple. Alternatively, we present the discovery reach  in terms of the suppression scale of the effective theory $\Lambda$ in the right panel. As can be seen, the current bounds allow for the detection of a sizable inelastic signal with a DARWIN-like detector. However, if elastic DM-nucleus scattering is not observed by XENONnT the room for a detectable inelastic signal shrinks substantially.

For the same experimental benchmark scenarios we show in Fig.~\ref{fig:DiscoveryReachOp} the discovery reach for the operators with a dominant elastic signal, i.e. $\mathcal{O}_5$, $\mathcal{O}_6$, $\mathcal{O}_9$ and $\mathcal{O}_{10}$, as a limit on the scale \footnote{The differential scattering cross section for these operators does not have the same properties as the one for SD-interactions and a recast of the bound as a limit on the zero momentum transfer cross section is highly misleading.}. The curve for the elastic reach plateaus  at $m_{\chi} \approx 200 \, \mbox{GeV}$ and remains fairly stable up to $1$ TeV. The inelastic reach is stronger than the elastic one at high mass for all operators and, therefore, a  clear detection of the inelastic signal in DARWIN is possible.  
 For lower masses the scale of new physics that can be probed in inelastic scattering  decreases rapidly and the competition with the elastic search channel becomes very challenging.

$\mathcal{O}_7$ and $\mathcal{O}_{13}$ lead to an inelastic scattering rate that exceeds the elastic one substantially. Depending on the DM mass, inelastic scattering could actually constitute the discovery channel for DM. Hence, we compare the discovery reach of both the elastic and the inelastic signal for a XENONnT-like detector and an exposure of $4$ tonne$\times$years in Fig.~\ref{fig:DiscoveryReachOp7p13}.  A search for an inelastic signal can easily outstrip the elastic search for $m_{\chi}\geq 60\, (50)$ GeV for $\mathcal{O}_{7}$ ($\mathcal{O}_{13}$). 
An analysis searching for inelastic scattering in already collected data could therefore produce the world's best limits on $\mathcal{O}_{7}$ and $\mathcal{O}_{13}$ or, more optimistically, lead to the first detection of DM in direct detection experiments.

\section{Discriminating DM nucleus interactions}
\label{sec:DiscriminationPower}

A DM signal in a direct detection experiment will be the starting point for numerous investigations into the properties of DM and its interactions with SM particles. Initially, the statistics of only a few events will not allow for a detailed analysis of the structure of DM interactions. However, more precise statements will become possible once more data are collected. 
Searching for inelastic DM-nucleus scattering is a potentially powerful way to exploit the data from ongoing experiments and learn more about DM.
In this section we will investigate whether the additional information encoded in inelastic scattering events can make a difference in our quest for the nature of DM and facilitate the identification of the true DM interaction. For related analyses without inelastic scattering see for example~\cite{Rogers:2016jrx,Fieguth:2018vob,Catena:2017wzu,Kahlhoefer:2016eds}.

We aim to assess whether it is possible to reject a default hypothesis in favor of the effective operator responsible for the interaction. To be more specific, we focus on interactions with inelastic scattering, take $\mathcal{O}_4$, i.e. the standard SD interaction, as our default hypothesis and determine the expected rejection power of this assumption in light of a hypothetical detection.

\subsection{Statistical treatment}

Given a signal generated by an operator $\mathcal{O}_B$ and an experimental background we aim to determine the statistical level with which we can reject the hypothesis that the signal is due to a different operator $\mathcal{O}_A$ as a function of experimental exposure. The operators will either describe only elastic scattering processes or additionally include inelastic interactions. By comparing the rejection power of these two hypotheses we can determine whether the inelastic operator contributions enhance the discrimination prospects. 

 We define two frequentist test statistics for hypotheses testing:
\begin{align}
 q_A:=-2\ln\left[\frac{\max\limits_{\Theta_{A}, \textbf{R}_{\textrm{BG}}}\mathscr{L}(\mathcal{D}_A|\Theta_{A},\textbf{R}_{\textrm{BG}})}{\max\limits_{\Theta_{B},\textbf{R}_{\textrm{BG}}}\mathscr{L}(\mathcal{D}_A|\Theta_{B},\textbf{R}_{\textrm{BG}})}\right]
  \quad\textrm{and}\quad 
  q_B:=-2\ln\left[\frac{\max\limits_{\Theta_{A}, \textbf{R}_{\textrm{BG}}}\mathscr{L}(\mathcal{D}_B|\Theta_{A},\textbf{R}_{\textrm{BG}})}{\max\limits_{\Theta_{B},\textbf{R}_{\textrm{BG}}}\mathscr{L}(\mathcal{D}_B|\Theta_{B},\textbf{R}_{\textrm{BG}})}\right],
\end{align}
where $\Theta_{i}:=(R_{i},m_{i})$ with $i\in\{A,B\}$. The likelihood function $\mathscr{L}$ remains as in Eq. (\ref{eq:LF}) with the exception that the DM mass is treated as a free parameter.
Both test statistics $q_A$ and $q_B$ test the hypothesis $\mathcal{H}_A$, which we take as null or default hypothesis ($A=\mathcal{O}_4$ in this work), against the alternative hypothesis $\mathcal{H}_B$ for different mock data. Each hypothesis is characterized by the binned probability distributions $f_{\mathcal{O}_{A/B},m_{A/B}}$ which are associated with the respective operators $\mathcal{O}_{A/B}$ and include either elastic or elastic plus inelastic effects. The two mock data sets $\mathcal{D}_{A/B}$ are generated from $f_{\mathcal{O}_{A/B},m_{A/B}}$ with fixed parameter values. A repetition of the calculation provides us with sets $\{q_A\}$ and $\{q_B\}$ from which we derive the distributions $f(q|\mathcal{D}_A)$ and $f(q|\mathcal{D}_B)$.
 The discrimination between different operators $\mathcal{O}_A$ and $\mathcal{O}_B$ requires the test statistics to be derived using mock data from both hypotheses since Wilk's and Wald's theorems apply in neither of them\footnote{In contrast to the rejection of the background-only hypothesis, the parameter spaces for the hypotheses testing of two distinct operators are not nested which violates a requirement of Wilk's theorem.}.

In detail our procedure is as follows: We start by simulating the data $\mathcal{D}_B(\overline{\mathcal{R}}_{B},\overline{m}_{B})$ from the background distributions using Poisson random sampling and $f_{\mathcal{O}_{B},m_{B}}$ with a fixed DM mass $\overline{m}_{B}=300\,\textrm{GeV}$ and a fixed DM rate $\overline{\mathcal{R}}_{B}=7\cdot 10^{-6} \,\mbox{kg}^{-1} \, \mbox{days}^{-1}$. Then we derive $q_B$ using the DM rate $\mathcal{R}_{B}$, the background rates $\textbf{R}_{\textrm{BG}}$ and the DM mass $m_{B}$ as fit parameters. Repeating this calculation $2500$ times enables us to derive the distribution $f(q|\mathcal{D}_B)$ and the median best fit mass $\hat{m}_B$ for the alternative hypothesis. The best fit mass $\hat{m}_B$ in turn is used as mass $\overline{m}_A$ for the production of $10000$ mock data $\mathcal{D}_A(\overline{\mathcal{R}}_{A},\overline{m}_{A})$ with $\overline{\mathcal{R}}_{A}=\overline{\mathcal{R}}_{B}$ which again are produced by Poisson random sampling the distribution over the bins. From that we find $\{q_A\}$ and in consequence $f(q|\mathcal{D}_A)$. This strategy is applied to our two cases in which we either consider only elastic scattering or include additionally the inelastic channels to both operators.

\begin{figure}[tb]
    \centering
        \includegraphics[width=0.49\textwidth]{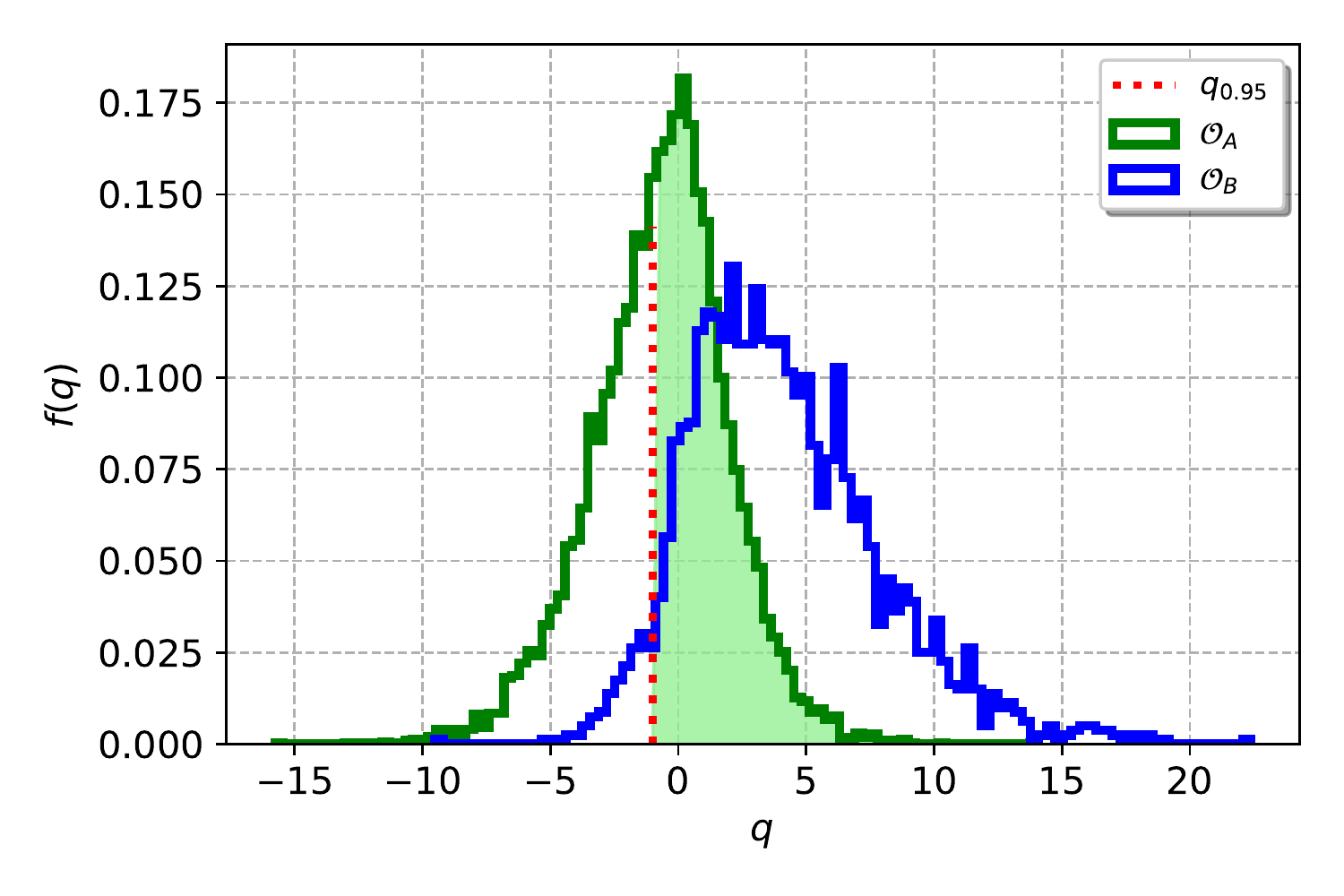} 
        \includegraphics[width=0.49\textwidth]{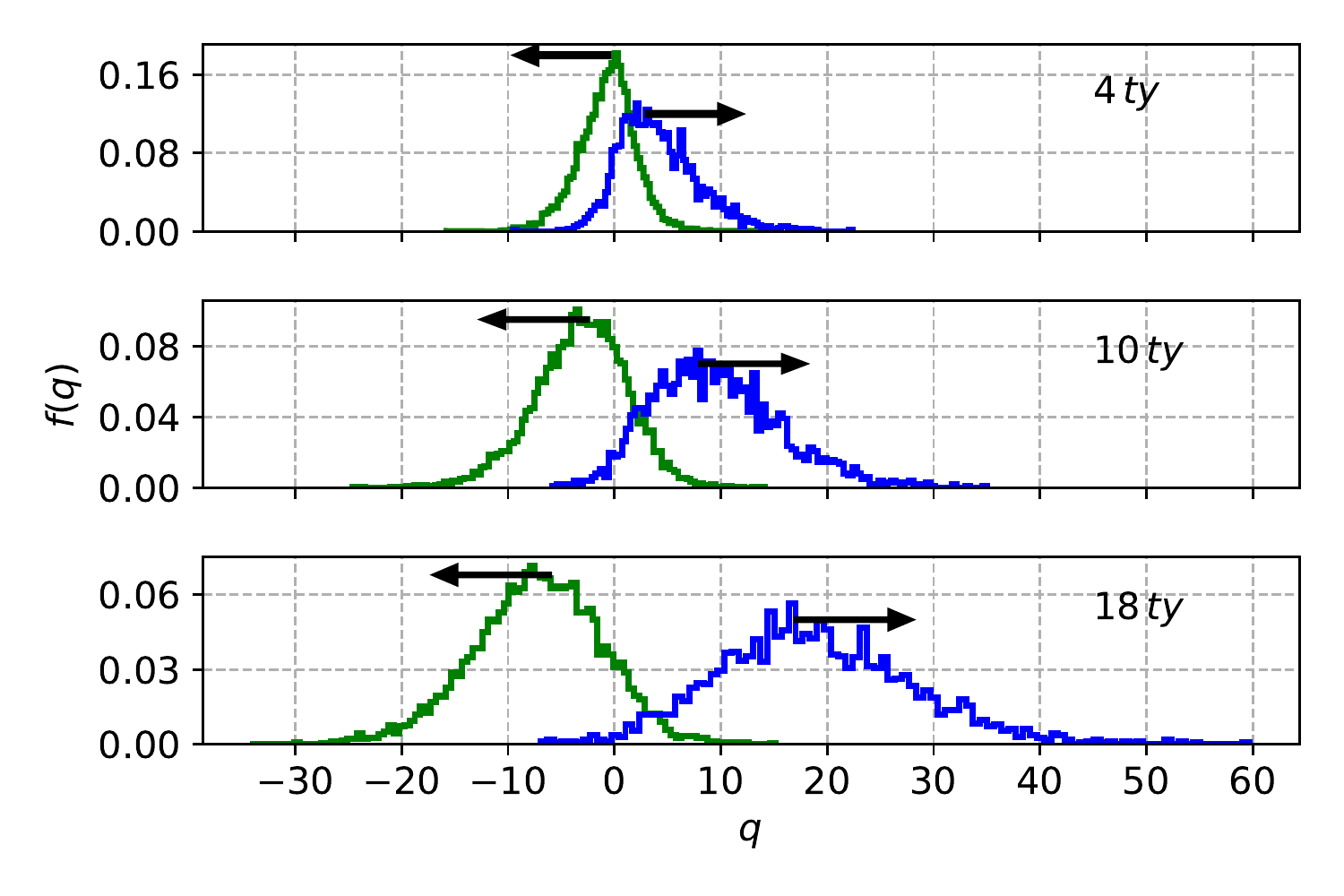}
    \caption{\textbf{Left:} As a generic example of $q$-value distributions for the hypothesis $\mathcal{H}_{\mathcal{O}_A}$  and the alternative hypothesis $\mathcal{H}_{\mathcal{O}_B}$ we show the distributions generated under the hypothesis $\mathcal{H}_{\mathcal{O}_4}$ (green histogram, $10000$ pseudo-experiments) and the alternative hypothesis $\mathcal{H}_{\mathcal{O}_5}$ (blue histogram, $2500$ pseudo-experiments) for an exposure of $4\,\textrm{t}\times\textrm{yr}$ and $m_{\chi}=1000\,\textrm{GeV}$. From the distribution of $\mathcal{H}_{\mathcal{O}_B}$ the $q_{0.95}$-value (red dotted line) is extracted which is used to calculate the $p$-value illustrated by the green area. \\\textbf{Right:} Impact of the exposure on the $q$-value distributions for the same hypotheses as in the left panel. For higher exposures signals become more distinct and the distributions move away from each other such that the $p$-value becomes smaller.}
    \label{fig:probdist}
\end{figure}

Our next goal is to derive the $p$-value as a measure of rejection. The hypothesis $\mathcal{H}_B$ can be rejected with probability $p$ against hypothesis $\mathcal{H}_A$ by
\begin{align}
 p:=\int_{q_{\textrm{z}}}^{\infty}f(q|\mathcal{D}_A)\,\textrm{d}q,
 \label{eq:power}
\end{align} 
where $q_{\textrm{z}}\in \{q_B\}$, such that $\int_{q_{\textrm{z}}}^{\infty}f({q|\mathcal{D}_B})\,\textrm{d}q=z$. We define the confidence level as $z=0.95$. In the left panel of Fig. \ref{fig:probdist} we illustrate the definition of the $p$-value for two example hypotheses. The right panel of the figure shows how the overlapping area underneath the two distributions $f(q|\mathcal{D}_A)$ and $f(q|\mathcal{D}_B)$ shrinks in consequence of an increasing exposure and thus leads to a decreasing $p$-value. 

As a result we find for each pair of DM operators the discrimination $p$-value as a function of the detector exposure for elastic only and elastic plus inelastic interactions. Finally, we compare the exposure values that allow to reject the hypothesis $\mathcal{H}_A$ with a statistical level of $2\,\sigma$ for the elastic only and elastic plus inelastic cases and deduce whether the inelastic signal contribution improves the discrimination between the two operators. 

\subsection{Results}

For the calculation of the discrimination power between two operators under the assumption of purely elastic or additional inelastic DM scattering we assume an elastic scattering rate of $\mathcal{R}_{\chi}=7.0\times10^{-6}\,\mbox{kg}^{-1} \, \mbox{days}^{-1}$ leading to $\approx 10$ events in the region of interest. This would allow for a $\approx 3 \sigma$ detection in an experiment with $4$ tonne$\times$years exposure assuming a DM mass of $m_{\chi}=300$~GeV as a representative example of heavy WIMP. We consider the six possibilities $\mathcal{H}_B \in \{ \mathcal{H}_{\mathcal{O}_5},\mathcal{H}_{\mathcal{O}_6},\mathcal{H}_{\mathcal{O}_7},\mathcal{H}_{\mathcal{O}_9},\mathcal{H}_{\mathcal{O}_{10}}, \mathcal{H}_{\mathcal{O}_{13}} \}$ while  $\mathcal{H}_A=\mathcal{H}_{\mathcal{O}_{4}}$. As observed in the previous section for the discovery reach, a clear separation between $\mathcal{O}_{7,13}$ and the other operators emerges.
With only $10$ events, elastic signals from different operators are basically indistinguishable and the discrimination $p$-values yield $\approx0.95$ and $\approx 0.66$ for $\mathcal{O}_7$ and $\mathcal{O}_{13}$, respectively. However, once the inelastic signal is considered the $p$-value drops well below $10^{-5}$.
This behavior is expected for $\mathcal{O}_7$ and $\mathcal{O}_{13}$ 
from the enhanced scattering rate in the inelastic channel w.r.t. the elastic channel. 

\begin{figure}[t]
    \centering
        \includegraphics[width=0.45\textwidth]{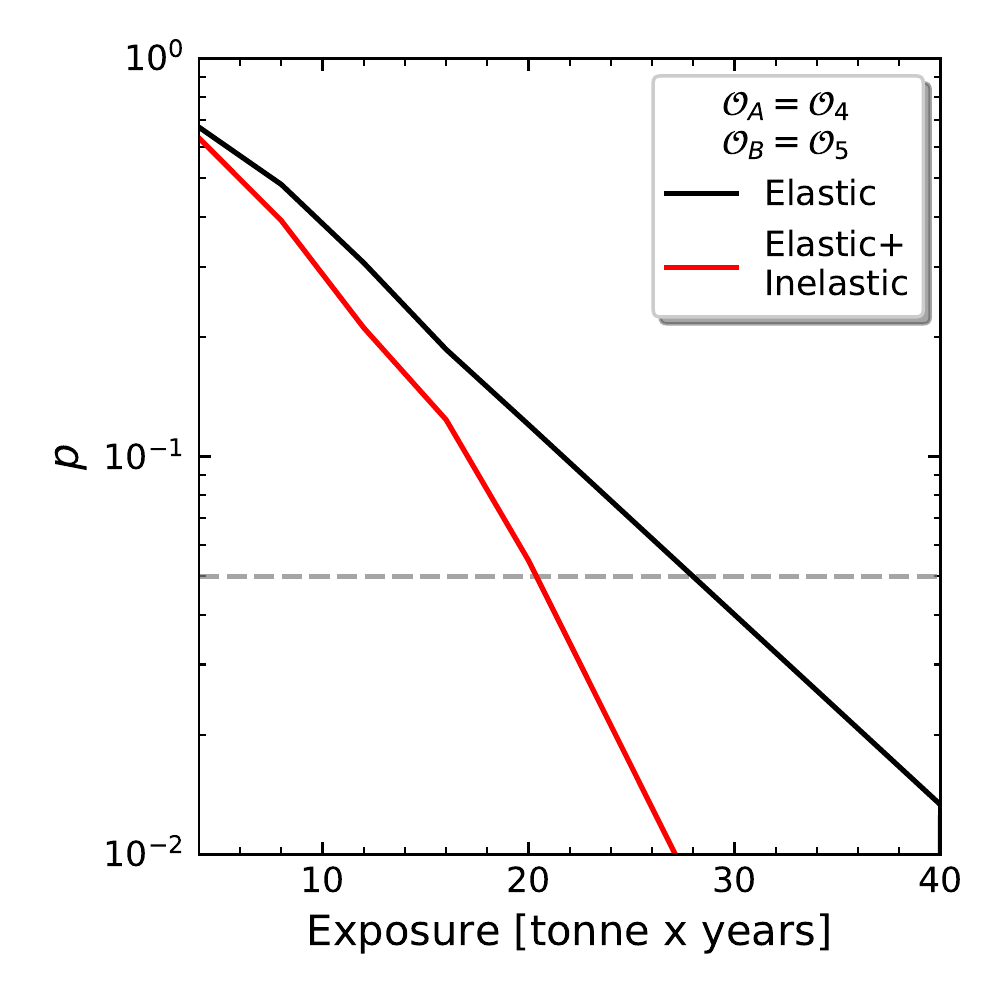} \qquad \includegraphics[width=0.45\textwidth]{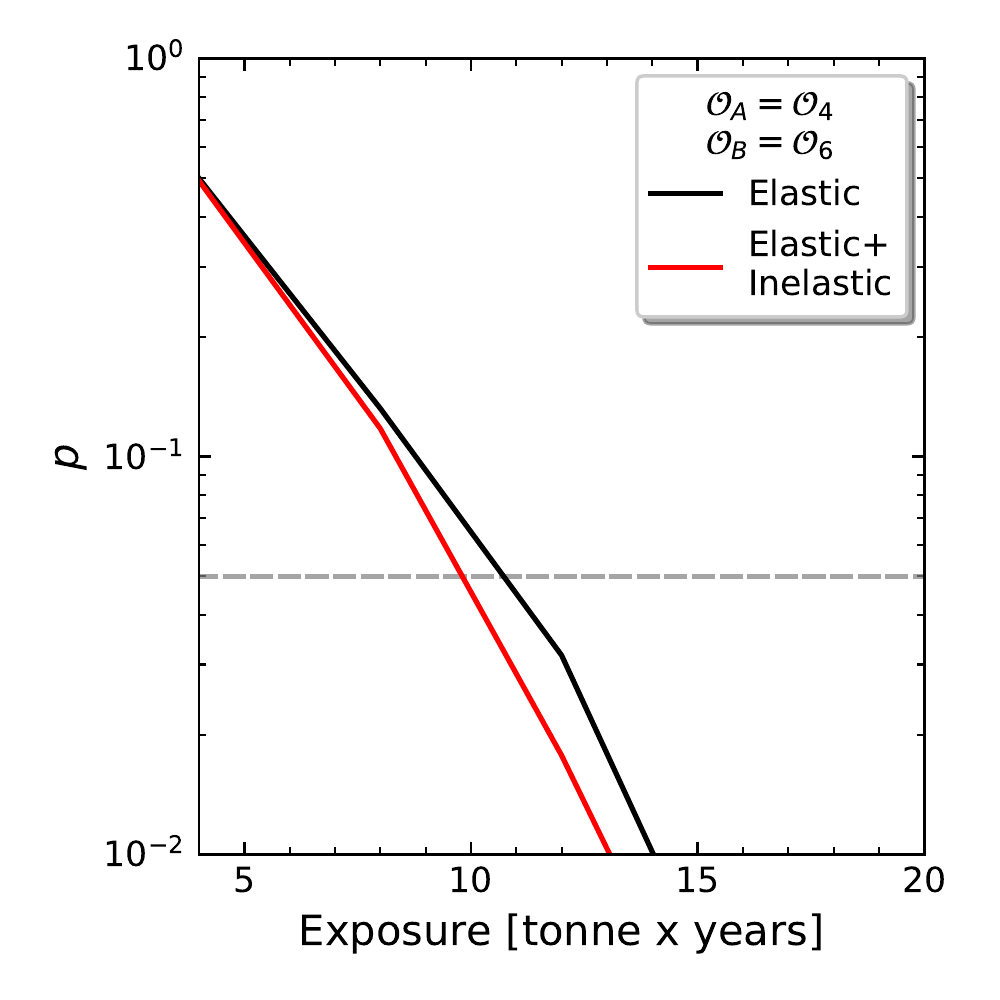}
        \includegraphics[width=0.45\textwidth]{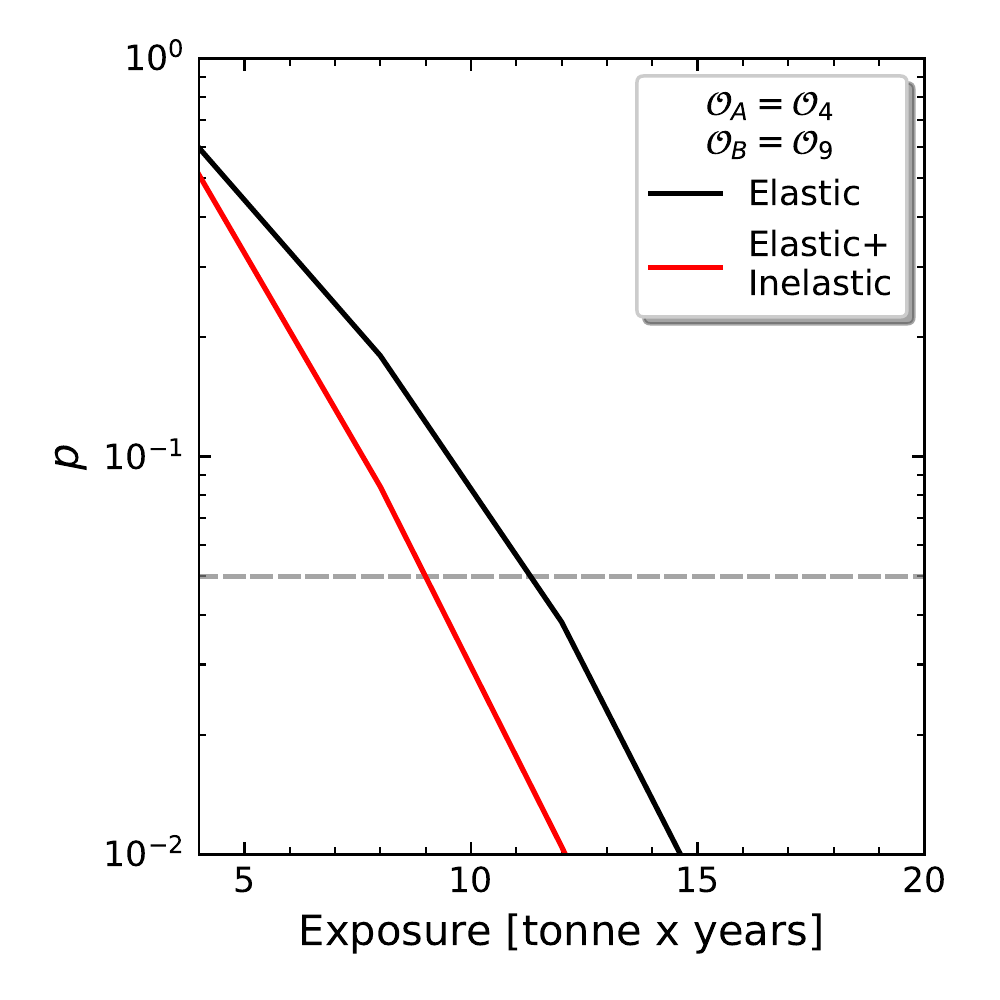}\qquad 
        \includegraphics[width=0.45\textwidth]{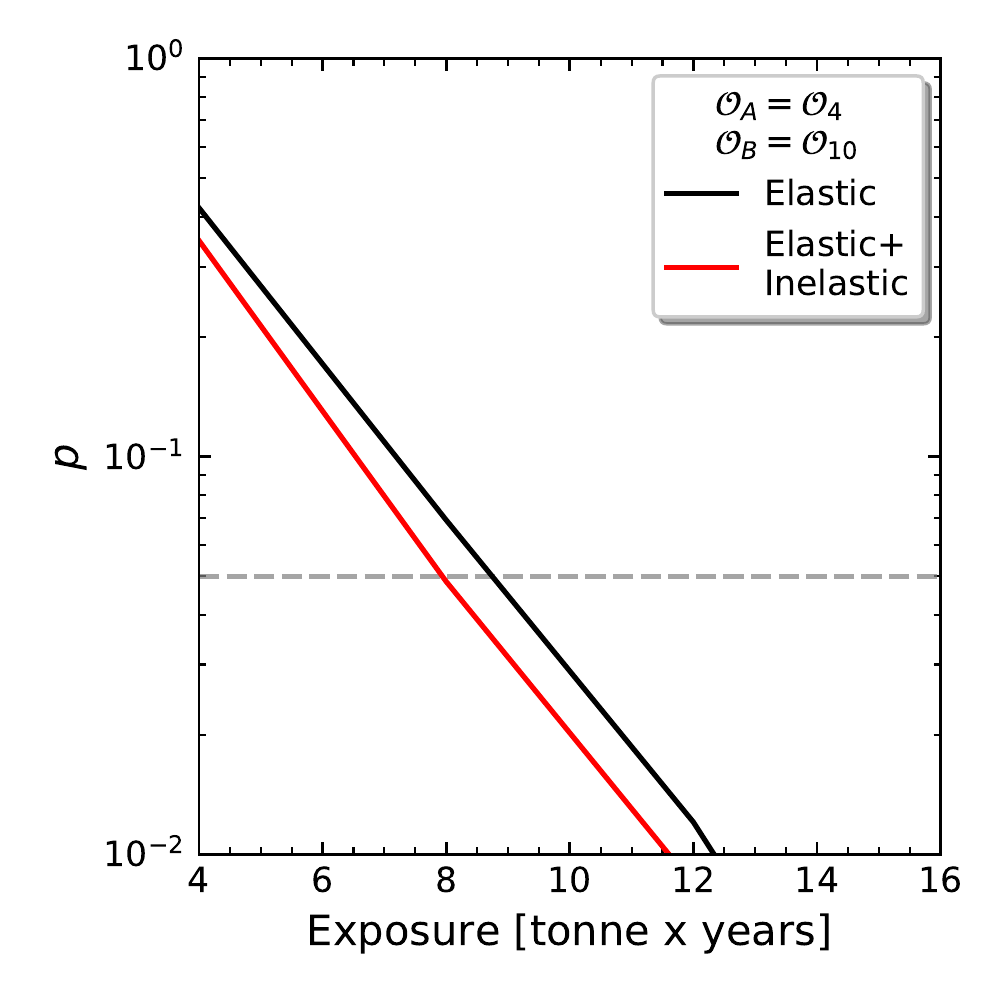}
    \caption{$p$-value as a function of exposure for $\mathcal{O}_5$ (top left), $\mathcal{O}_6$ (top right), $\mathcal{O}_9$ (bottom left) and $\mathcal{O}_{10}$ (bottom right) for an elastic scattering rate of $7 \times 10^{-5}\, \mbox{kg}^{-1} \; \mbox{days}^{-1}$ and $m_{\chi}= 300\,$~GeV. Results for an elastic plus inelastic  signal are shown in red while results of analyses utilizing only the elastic signal are shown in black. The gray dashed line indicates $p=0.05$. }
    \label{fig:OpDiscrimination}
\end{figure}

For the other four operators the inelastic signal rate is suppressed compared to the elastic one. $10$ elastic events correspond to $\approx 1$ inelastic scattering events. Therefore, the inelastic and the elastic signal lead to the same discrimination $p$-values of $\mathcal{O}(0.1)$ for this chosen exposure. In Fig.~\ref{fig:OpDiscrimination} we show the evolution of the $p$-values calculated for purely elastic DM scattering or an additional inelastic interaction as a function of exposure.
The inelastic signal substantially improves the $p$-value for $\mathcal{O}_5$ (top left) and the exposure required to exclude the default hypothesis at the $p=0.05$ level, $E_{p=0.05}$, decreases by more than a factor of 2. $E_{p=0.05}$  is in reach of an extended run of the discovery device while reaching the same level of confidence only based on the elastic signal would probably require to scale up the detector's target mass.

In the case of $\mathcal{O}_{9}$ the consideration of inelastic signals decreases $E_{p=0.05}$ from $\approx  11.5\; \mbox{tonne}\times\mbox{years}$ to $9\; \mbox{tonne}\times\mbox{years}$.
For $\mathcal{O}_{6}$ and $\mathcal{O}_{10}$, the discrimination $p$-values in the vicinity of $E_{p=0.05}$ for the elastic-only case are $\approx 50\%$ weaker than those from the elastic plus inelastic channel.

Summarizing our findings, inelastic scattering events improve the discrimination power for four of the six considered operators and add a moderate improvement for the other two. Therefore, a search for inelastic scattering is going to provide valuable additional information. Given that the required data are acquired simultaneously to the elastic scattering data, the inelastic channel offers a potential to improve our understanding of DM without constructing additional experiments.  

\section{Conclusion}
\label{sec:Conclusions}

Direct detection experiments offer one of the most promising ways to search for DM. In light of the substantial gains in sensitivity of future experiments, a detection of the DM particle is an exciting possibility.
Once the detection has been established, the next question will be what about the DM particle's nature that can be extracted from the observations. Subleading signatures, which will become accessible once a detection has been established, might play a major role in answering this question. 
Direct detection experiments are not only sensitive to the standard signature, i.e. the energy deposit for a nucleus recoiling against the DM in an elastic scattering process, but can also probe inelastic scattering in which the nucleus is excited.  
Xenon-based direct detection experiments are particularly interesting in this context since they have shown leading sensitivities to DM and the two common isotopes $^{129}$Xe and $^{131}$Xe have low-lying first excited states. In this paper we quantified the physics reach of future xenon-detectors w.r.t. the full set of non-relativistic operators. We extend the non-relativistic effective theory to account for inelastic processes and perform a nuclear shell model calculation for the ground and the first excited state. Combined with a full Monte-Carlo simulation of a realistic two-phase xenon-detector this allows us to derive the discovery reach for inelastic scattering. 

We find encouraging prospects for the detection of inelastic scattering with future xenon-based detectors for eight of the $14$ non-relativistic effective operators. In the case of two operators, $\mathcal{O}_{7,13}$, the prospects for discovery of DM induced nuclear excitations are even better than for the discovery of the conventional elastic signal. Consequently, an analysis of current data with inelastic scattering in mind would lead to the world's best limits on these operators or might even reveal the first direct detection of DM.   The additional information encoded in the inelastic scattering events can be used to learn more about the theory of DM once a signal is established. We have analyzed the power with which the standard hypothesis of DM interactions with inelastic interactions, i.e. the standard spin-dependent interaction $\mathcal{O}_4$, can be rejected if the data follows one of the other operators that lead to inelastic scattering. We found that including the inelastic signal improves the $p$-value significantly in four of the six cases under consideration. In the other cases the elastic signal itself is already distinct enough to shift the $p$-value to $0.05$ rather fast with increasing exposure and the benefit of using inelastic scattering events is less prominent. 

In light of the excellent prospects for $\mathcal{O}_{7,13}$ we would like to encourage the experimental community to revisit the collected data with inelastic scattering in mind and perform a dedicated analysis. We also want to stress the importance of using all available data once a DM signal is discovered.

\section*{Acknowledgements}

We thank  M. Lindner, T. Marrodan Undagoitia and C. McCabe for helpful discussions about xenon-based dark matter detectors in general and inelastic scattering in particular. We also thank B. A. Brown for providing the NushellX@MSU code for shell model calculations. 

\begin{appendix}
\section{Detector simulation}
\label{sec:app_simulation}

In this section we give further details on the parameters employed in the detector simulation that converts energy deposition in LXe into the measured signals S1 and S2$_\mathrm{b}$. A list of the parameters and how they are treated for ERs and NRs is given in Tab. \ref{tab:LxeParams}:

\begin{itemize}
    \item The Lindhard quenching factor $L$ quantifies the fraction of the energy that is converted into visible signals, i.e. light or ionization charges. For ERs this fraction is assumed to be 100\% while part of the energy is lost to atomic motion for NRs. The parameter $k$ in the formula give in Tab. \ref{tab:LxeParams} relates the electronic stopping power and the velocity of NRs and yields 0.1394. The function $g(\kappa)$ is parametrized by
\begin{equation}
    g = 3\,\kappa^{0.15}+0.7\,\kappa^{0.6}+\kappa,
\end{equation}
with $\kappa = 11.5\, E\cdot Z^{-7/3}$ that depends on the electric field $E$ and the proton number $Z$, i.e. 54 for xenon.
\item The Fano factor $F$ is a measure for the fluctuation of the number of ions and scintillation photons produced by an energy deposition and has been measured to yield a value of 0.03 for ionizing radiation. Since there are no dedicated measurements for $F$ in LXe for NRs, we assume the same value for NRs as for ERs.
\item The ratio $n_\mathrm{ex}/n_\mathrm{i}$ of the number of excitons to ions created in the interaction is fixed for ERs while it follows a field dependent function for NRs (see Tab. \ref{tab:LxeParams}).
\item The probability $r$ for an electron-ion pair to recombine after its creating is determined by the Thomas-Imel box model~\cite{ThomasImel}. For NRs it takes the parameters $\xi = \gamma\cdot E^{-\delta}$, $\gamma = 0.014$ and $\delta = 0.062$ which have been estimated from experimental data. The expression for $\xi$ in the case of ERs is more complicated and depends on both, the deposited energy $\epsilon$ and $E$. 
For simplicity, $n_\gamma$ and $n_\mathrm{e}$, i.e. the mean number of photons and electrons emitted and extracted from the interaction site, are obtained from the average light and charge yield curves for ERs given in~\cite{NEST_2013}. The numbers are used to re-calculated the recombination fraction. During this work the new version NESTv2 became available~\cite{NESTv2}. The predicted light and charge yield values agree with the previous version within 20\%.
\item The standard deviation of recombination fluctuations $\sigma_\mathrm{r}$ yields a constant value of 0.06 for ERs while for NRs it has been shown to be dependent on $r$.
\item In the case of NRs two excitons can interact to produce a single photon leading to an additional quenching factor $f_\text{l}$ (Penning quenching) that is parametrized by $\kappa$ (see Lindhard quenching), $\eta = 3.32$ and $\lambda = 1.14$.
\end{itemize}

\renewcommand{\arraystretch}{1.5}
\begin{table}
 \begin{tabular}{ccccc}
 Description & Parameter &  ER & NR & Reference \\\hline
 Lindhard quenching factor & $L$ & 1 & $\frac{k\cdot g(\kappa)}{1+k\cdot g(\kappa)}$ & \cite{NEST_2015}\\
 Fano factor & $F$ & 0.03 & 0.03 & \cite{FanoFactor_1995}\\
 Exciton-ion-ratio & $n_\mathrm{ex}/n_\mathrm{i}$ & 0.15 & $1.240\cdot E^{-0.0472}\cdot(1-e^{-239\kappa})$ &  \cite{NEST_2015}\\
 Recombination probability & $r$ & $ \frac{(n_\gamma/n_\mathrm{e})-n_\mathrm{ex}/n_\mathrm{e}}{(n_\gamma/n_\mathrm{e})+1}$ & $1-\frac{\mathrm{ln}(1+n_\mathrm{i}\,\xi)}{n_\mathrm{i}\,\xi}$& \cite{ThomasImel,NEST_2013}\\
 Recombination fluctuation & $\sigma_\mathrm{r}$ & 0.06 & $n_\mathrm{i}\sqrt{0.0056\,(1-r)}$ & \cite{NEST_2015}\\
 Penning quenching & $f_\text{l}$ & 1 & $\frac{1}{1+\eta \kappa^\lambda}$ & \cite{NEST_2015}
 \end{tabular}
 \caption{List of parameters used in the simulation of liquid xenon microphysics. See text for further explanations.}
 \label{tab:LxeParams}
\end{table}
\renewcommand{\arraystretch}{1}


\section{DM response function}
\label{Appx:DMresponse}
\begin{align*}
    R_{M}^{\tau \tau'}\left( v^{\perp 2}_T,\frac{q^2}{m_N^2} \right) =& c_1^\tau c_1^{\tau'} + \frac{J_\chi(J_\chi+1)}{3} \left( \frac{q^2}{m_N^2} v^{\perp 2}_T c_5^\tau c_5^{\tau'} + v^{\perp 2} c_8^\tau c_8^{\tau'} + \frac{q^2}{m_N^2}c_{11}^\tau c_{11}^{\tau'}\right)\\
    R_{\Phi''}^{\tau \tau'}\left( v^{\perp 2}_T,\frac{q^2}{m_N^2} \right) =&  \frac{q^2}{4 m_N^2}c_{3}^\tau c_{3}^{\tau'} +  \frac{J_\chi(J_\chi+1)}{12} \left( c_{12}^\tau - \frac{q^2}{m_N^2} c_{15}^{\tau}\right)\left( c_{12}^{\tau'} - \frac{q^2}{m_N^2} c_{15}^{\tau'}\right)\\
    R_{\Phi''M}^{\tau \tau'}\left( v^{\perp 2}_T,\frac{q^2}{m_N^2} \right) =&  c_{3}^\tau c_{1}^{\tau'} +  \frac{J_\chi(J_\chi+1)}{3} \left( c_{12}^\tau - \frac{q^2}{m_N^2} c_{15}^{\tau}\right)c_{11}^{\tau'}\\
     R_{\tilde{\Phi}'}^{\tau \tau'}\left( v^{\perp 2}_T,\frac{q^2}{m_N^2} \right) =& \frac{J_\chi (J_\chi+1)}{12}\left(c_{12}^\tau c_{12}^{\tau'} + \frac{q^2}{m_N^2} c_{13}^\tau c_{13}^{\tau'} \right) \\
     R_{\Sigma''}^{\tau \tau'}\left( v^{\perp 2}_T,\frac{q^2}{m_N^2} \right) =& \frac{q^2}{4 m_N^2} c_{10}^\tau c_{10}^{\tau'} + \frac{J_\chi (J_\chi+1)}{12} \left( c_4^\tau c_4^{\tau'} +  \right. \\
     &\left. \frac{q^2}{m_N^2}(c_4^\tau c_6^{\tau'} + c_4^{\tau'} c_6^{\tau}) + \frac{q^4}{m_N^2} c_6^{\tau} c_6^{\tau'} +  v^{\perp 2}_T c_{12}^{\tau} c_{12}^{\tau'} + \frac{q^2}{m_N^2} v^{\perp 2}_T c_{13}^\tau c_{13}^{\tau'}  \right)\\
     R_{\Sigma'}^{\tau \tau'}\left( v^{\perp 2}_T,\frac{q^2}{m_N^2} \right) =& \frac{1}{8}\left(  \frac{q^2}{m_N^2} v_T^{\perp 2} c_{3}^\tau c_{3}^{\tau'} + v_T^{\perp 2} c_7^\tau c_7^{\tau'} \right) +  \frac{J_\chi (J_\chi+1)}{12}\left[ c_4^{\tau} c_4^{\tau'} \right.\\
     & \left. \frac{q^2}{m_N^2}c_9^\tau c_9^{\tau'} + \frac{v_T^{\perp 2}}{2} \left( c_{12}^{\tau} - \frac{q^2}{m_N^2}c_{15}^{\tau}\right) \left( c_{12}^{\tau'} - \frac{q^2}{m_N^2}c_{15}^{\tau'}\right)  + \frac{q^2}{2 m_N^2} v^{\perp 2}_T c_{14}^\tau c_{14}^{\tau'}\right]\\
      R_{\Delta}^{\tau \tau'}\left( v^{\perp 2}_T,\frac{q^2}{m_N^2} \right) =& \frac{J_\chi (J_\chi +1)}{3} \left( \frac{q^2}{m_N^2} c_5^\tau c_5^{\tau'} + c_8^\tau c_8^{\tau'} \right)\\
      R_{\Delta \Sigma'}^{\tau \tau'}\left( v^{\perp 2}_T,\frac{q^2}{m_N^2} \right) =& \frac{J_\chi (J_\chi +1)}{3} \left( c_5^\tau c_4^{\tau'} -c_8^\tau c_9^{\tau'} \right)\\
       R_{\Delta'}^{\tau \tau'}\left( v^{\perp 2}_T,\frac{q^2}{m_N^2} \right) =& \frac{J_\chi (J_\chi +1)}{3} \left( \frac{q^2}{m_N^2} c_5^\tau c_5^{\tau'} + c_8^\tau c_8^{\tau'} \right)\\
       \end{align*}
       \begin{align*}
       R_{\Sigma}^{\tau \tau'}\left( v^{\perp 2}_T,\frac{q^2}{m_N^2} \right) =& \frac{1}{8}\left(  \frac{q^2}{m_N^2} v_T^{\perp 2} c_{3}^\tau c_{3}^{\tau'} + v_T^{\perp 2} c_7^\tau c_7^{\tau'} \right) +  \frac{J_\chi (J_\chi+1)}{12}\left[ c_4^{\tau} c_4^{\tau'} \right.\\
        & \left. \frac{q^2}{m_N^2}c_9^\tau c_9^{\tau'} + \frac{v_T^{\perp 2}}{2} \left( c_{12}^{\tau} - \frac{q^2}{m_N^2}c_{15}^{\tau}\right) \left( c_{12}^{\tau'} - \frac{q^2}{m_N^2}c_{15}^{\tau'}\right)  + \frac{q^2}{2 m_N^2} v^{\perp 2}_T c_{14}^\tau c_{14}^{\tau'}\right]\\
       R_{\tilde{\Phi}}^{\tau \tau'}\left( v^{\perp 2}_T,\frac{q^2}{m_N^2} \right) =& \frac{J_\chi (J_\chi+1)}{12}\left(c_{12}^\tau c_{12}^{\tau'} + \frac{q^2}{m_N^2} c_{13}^\tau c_{13}^{\tau'} \right) \\
       R_{\tilde{\Omega}}^{\tau \tau'}\left( v^{\perp 2}_T,\frac{q^2}{m_N^2} \right) =& \frac{q^2}{4 m_N^2}\left(c_7^\tau c_7^{\tau'} + \frac{J_\chi (J_\chi +1)}{3} \frac{q^2}{m_N^2} c_{14}^\tau c_{14}^{\tau'}\right)\\
       R_{\tilde{\Delta}''}^{\tau \tau'}\left( v^{\perp 2}_T,\frac{q^2}{m_N^2} \right) =& \frac{q^2}{m_N^2} \frac{J_\chi (J_\chi+1)}{3} c_8^\tau c_8^{\tau'}\\
       R_{\Sigma \Delta'}^{\tau \tau'}\left( v^{\perp 2}_T,\frac{q^2}{m_N^2} \right) =& \frac{J_\chi (J_\chi +1)}{3} \left( c_5^\tau c_4^{\tau'} -c_8^\tau c_9^{\tau'} \right)\\
\end{align*}

\end{appendix}
\bibliographystyle{JHEPfixed}
\bibliography{bibliography}

\providecommand{\href}[2]{#2}\begingroup\raggedright\begin{thebibliography}{10}

\bibitem{Aprile:2013doa}
{\bf XENON100} Collaboration, E.~Aprile {\em et.~al.}, {\it {Limits on
  spin-dependent WIMP-nucleon cross sections from 225 live days of XENON100
  data}},  {\em Phys. Rev. Lett.} {\bf 111} (2013), no.~2 021301,
  [\href{http://xxx.lanl.gov/abs/1301.6620}{{\tt 1301.6620}}].

\bibitem{Xe1T-SI}
{\bf XENON} Collaboration, E.~Aprile {\em et.~al.}, {\it Dark matter search
  results from a one ton-year exposure of xenon1t},  {\em Phys. Rev. Lett.}
  {\bf 121} (Sep, 2018) 111302.

\bibitem{LUX-SI}
{\bf LUX} Collaboration, D.~S. Akerib {\em et.~al.}, {\it Results from a search
  for dark matter in the complete lux exposure},  {\em Phys. Rev. Lett.} {\bf
  118} (Jan, 2017) 021303.

\bibitem{PandaX-SI}
{\bf PandaX} Collaboration, X.~Cui {\em et.~al.}, {\it Dark matter results from
  54-ton-day exposure of pandax-ii experiment},  {\em Phys. Rev. Lett.} {\bf
  119} (Oct, 2017) 181302.

\bibitem{Fitzpatrick:2012ix}
A.~L. Fitzpatrick, W.~Haxton, E.~Katz, N.~Lubbers, and Y.~Xu, {\it {The
  Effective Field Theory of Dark Matter Direct Detection}},  {\em JCAP} {\bf
  1302} (2013) 004, [\href{http://xxx.lanl.gov/abs/1203.3542}{{\tt
  1203.3542}}].

\bibitem{DelNobile:2013sia}
M.~Cirelli, E.~Del~Nobile, and P.~Panci, {\it {Tools for model-independent
  bounds in direct dark matter searches}},  {\em JCAP} {\bf 1310} (2013) 019,
  [\href{http://xxx.lanl.gov/abs/1307.5955}{{\tt 1307.5955}}].

\bibitem{Fitzpatrick:2012ib}
A.~L. Fitzpatrick, W.~Haxton, E.~Katz, N.~Lubbers, and Y.~Xu, {\it {Model
  Independent Direct Detection Analyses}},
  \href{http://xxx.lanl.gov/abs/1211.2818}{{\tt 1211.2818}}.

\bibitem{Catena:2014uqa}
R.~Catena and P.~Gondolo, {\it {Global fits of the dark matter-nucleon
  effective interactions}},  {\em JCAP} {\bf 1409} (2014), no.~09 045,
  [\href{http://xxx.lanl.gov/abs/1405.2637}{{\tt 1405.2637}}].

\bibitem{Catena:2015uha}
R.~Catena and B.~Schwabe, {\it {Form factors for dark matter capture by the Sun
  in effective theories}},  {\em JCAP} {\bf 1504} (2015), no.~04 042,
  [\href{http://xxx.lanl.gov/abs/1501.03729}{{\tt 1501.03729}}].

\bibitem{Catena:2016hoj}
R.~Catena, A.~Ibarra, and S.~Wild, {\it {DAMA confronts null searches in the
  effective theory of dark matter-nucleon interactions}},  {\em JCAP} {\bf
  1605} (2016), no.~05 039, [\href{http://xxx.lanl.gov/abs/1602.04074}{{\tt
  1602.04074}}].

\bibitem{Dent:2015zpa}
J.~B. Dent, L.~M. Krauss, J.~L. Newstead, and S.~Sabharwal, {\it {General
  analysis of direct dark matter detection: From microphysics to observational
  signatures}},  {\em Phys. Rev.} {\bf D92} (2015), no.~6 063515,
  [\href{http://xxx.lanl.gov/abs/1505.03117}{{\tt 1505.03117}}].

\bibitem{Gresham:2014vja}
M.~I. Gresham and K.~M. Zurek, {\it {Effect of nuclear response functions in
  dark matter direct detection}},  {\em Phys. Rev.} {\bf D89} (2014), no.~12
  123521, [\href{http://xxx.lanl.gov/abs/1401.3739}{{\tt 1401.3739}}].

\bibitem{Boddy:2018kfv}
K.~K. Boddy and V.~Gluscevic, {\it {First Cosmological Constraint on the
  Effective Theory of Dark Matter-Proton Interactions}},  {\em Phys. Rev.} {\bf
  D98} (2018), no.~8 083510, [\href{http://xxx.lanl.gov/abs/1801.08609}{{\tt
  1801.08609}}].

\bibitem{Kavanagh:2015jma}
B.~J. Kavanagh, {\it {New directional signatures from the nonrelativistic
  effective field theory of dark matter}},  {\em Phys. Rev.} {\bf D92} (2015),
  no.~2 023513, [\href{http://xxx.lanl.gov/abs/1505.07406}{{\tt 1505.07406}}].

\bibitem{Aprile:2017aas}
{\bf XENON} Collaboration, E.~Aprile {\em et.~al.}, {\it {Effective field
  theory search for high-energy nuclear recoils using the XENON100 dark matter
  detector}},  {\em Phys. Rev.} {\bf D96} (2017), no.~4 042004,
  [\href{http://xxx.lanl.gov/abs/1705.02614}{{\tt 1705.02614}}].

\bibitem{Angloher:2018fcs}
G.~Angloher {\em et.~al.}, {\it {Limits on Dark Matter Effective Field Theory
  Parameters with CRESST-II}},  \href{http://xxx.lanl.gov/abs/1809.03753}{{\tt
  1809.03753}}.

\bibitem{Goodman:1984dc}
M.~W. Goodman and E.~Witten, {\it {Detectability of Certain Dark Matter
  Candidates}},  {\em Phys. Rev.} {\bf D31} (1985) 3059. [,325(1984)].

\bibitem{Ellis:1988nb}
J.~R. Ellis, R.~A. Flores, and J.~D. Lewin, {\it {Rates for Inelastic Nuclear
  Excitation by Dark Matter Particles}},  {\em Phys. Lett.} {\bf B212} (1988)
  375--380.

\bibitem{Engel:1999kv}
J.~Engel and P.~Vogel, {\it {Neutralino inelastic scattering with subsequent
  detection of nuclear gamma-rays}},  {\em Phys. Rev.} {\bf D61} (2000) 063503,
  [\href{http://xxx.lanl.gov/abs/hep-ph/9910409}{{\tt hep-ph/9910409}}].

\bibitem{Xe1T-Exp}
{\bf XENON} Collaboration, E.~Aprile {\em et.~al.}, {\it The xenon1t dark
  matter experiment},  {\em The European Physical Journal C} {\bf 77} (Dec,
  2017) 881.

\bibitem{LUX-Exp}
{\bf LUX} Collaboration, D.~Akerib {\em et.~al.}, {\it The large underground
  xenon (lux) experiment},  {\em Nuclear Instruments and Methods in Physics
  Research Section A: Accelerators, Spectrometers, Detectors and Associated
  Equipment} {\bf 704} (2013) 111 -- 126.

\bibitem{PandaX-Exp}
{\bf PandaX} Collaboration, X.~Cao {\em et.~al.}, {\it Pandax: a liquid xenon
  dark matter experiment at cjpl},  {\em Science China Physics, Mechanics {\&}
  Astronomy} {\bf 57} (Aug, 2014) 1476--1494.

\bibitem{Aprile:2017ngb}
{\bf XENON} Collaboration, E.~Aprile {\em et.~al.}, {\it {Search for WIMP
  Inelastic Scattering off Xenon Nuclei with XENON100}},  {\em Phys. Rev.} {\bf
  D96} (2017), no.~2 022008, [\href{http://xxx.lanl.gov/abs/1705.05830}{{\tt
  1705.05830}}].

\bibitem{PandaX-Inelastic}
{\bf PandaX} Collaboration, X.~Chen {\em et.~al.}, {\it Exploring the dark
  matter inelastic frontier with 79.6 days of pandax-ii data},  {\em Phys. Rev.
  D} {\bf 96} (Nov, 2017) 102007.

\bibitem{Suzuki:2018xek}
{\bf XMASS} Collaboration, T.~Suzuki {\em et.~al.}, {\it {Search for
  WIMP-$^{129}$Xe inelastic scattering with particle identification in
  XMASS-I}},  \href{http://xxx.lanl.gov/abs/1809.05358}{{\tt 1809.05358}}.

\bibitem{Baudis:2013bba}
L.~Baudis, G.~Kessler, P.~Klos, R.~F. Lang, J.~Menendez, S.~Reichard, and
  A.~Schwenk, {\it {Signatures of Dark Matter Scattering Inelastically Off
  Nuclei}},  {\em Phys. Rev.} {\bf D88} (2013), no.~11 115014,
  [\href{http://xxx.lanl.gov/abs/1309.0825}{{\tt 1309.0825}}].

\bibitem{McCabe:2015eia}
C.~McCabe, {\it {Prospects for dark matter detection with inelastic transitions
  of xenon}},  {\em JCAP} {\bf 1605} (2016), no.~05 033,
  [\href{http://xxx.lanl.gov/abs/1512.00460}{{\tt 1512.00460}}].

\bibitem{DARWIN}
{\bf DARWIN} Collaboration, J.~Aalbers {\em et.~al.}, {\it {DARWIN: towards the
  ultimate dark matter detector}},  {\em JCAP} {\bf 1611} (2016) 017,
  [\href{http://xxx.lanl.gov/abs/1606.07001}{{\tt 1606.07001}}].

\bibitem{Bishara:2017pfq}
F.~Bishara, J.~Brod, B.~Grinstein, and J.~Zupan, {\it {From quarks to nucleons
  in dark matter direct detection}},  {\em JHEP} {\bf 11} (2017) 059,
  [\href{http://xxx.lanl.gov/abs/1707.06998}{{\tt 1707.06998}}].

\bibitem{Hoferichter:2018acd}
M.~Hoferichter, P.~Klos, J.~Menéndez, and A.~Schwenk, {\it {Nuclear structure
  factors for general spin-independent WIMP-nucleus scattering}},  {\em Phys.
  Rev.} {\bf D99} (2019), no.~5 055031,
  [\href{http://xxx.lanl.gov/abs/1812.05617}{{\tt 1812.05617}}].

\bibitem{Anand:2013yka}
N.~Anand, A.~L. Fitzpatrick, and W.~C. Haxton, {\it {Weakly interacting massive
  particle-nucleus elastic scattering response}},  {\em Phys. Rev.} {\bf C89}
  (2014), no.~6 065501, [\href{http://xxx.lanl.gov/abs/1308.6288}{{\tt
  1308.6288}}].

\bibitem{dmformfactor}
\url{https://www.ocf.berkeley.edu/~nanand/software/dmformfactor/}.

\bibitem{Catena:2019hzw}
R.~Catena, K.~Fridell, and M.~B. Krauss, {\it {Non-relativistic Effective
  Interactions of Spin 1 Dark Matter}},  {\em JHEP} {\bf 08} (2019) 030,
  [\href{http://xxx.lanl.gov/abs/1907.02910}{{\tt 1907.02910}}].

\bibitem{Donnelly:1978tz}
T.~W. Donnelly and R.~D. Peccei, {\it {Neutral Current Effects in Nuclei}},
  {\em Phys. Rept.} {\bf 50} (1979) 1.

\bibitem{Walecka:1995mi}
J.~D. Walecka, {\it {Theoretical nuclear and subnuclear physics}},  {\em Oxford
  Stud. Nucl. Phys.} {\bf 16} (1995) 1--610.

\bibitem{Serot:1978vj}
B.~D. Serot, {\it {Semileptonic Weak and Electromagnetic Interactions with
  Nuclei: Nuclear Current Operators Through Order (v/c)**2 (Nucleon)}},  {\em
  Nucl. Phys.} {\bf A308} (1978) 457--499.

\bibitem{Walecka:2001gs}
J.~D. Walecka, {\em {Electron scattering for nuclear and nucleon structure}},
  vol.~16.
\newblock Cambridge University Press, 2005.

\bibitem{Brown:2014bhl}
B.~A. Brown and W.~D.~M. Rae, {\it {The Shell-Model Code NuShellX@MSU}},  {\em
  Nucl. Data Sheets} {\bf 120} (2014) 115--118.

\bibitem{Brown:2004xk}
B.~A. Brown, N.~J. Stone, J.~R. Stone, I.~S. Towner, and M.~Hjorth-Jensen, {\it
  {Magnetic moments of the 2+(1) states around Sn-132}},  {\em Phys. Rev.} {\bf
  C71} (2005) 044317, [\href{http://xxx.lanl.gov/abs/nucl-th/0411099}{{\tt
  nucl-th/0411099}}]. [Erratum: Phys. Rev.C72,029901(2005)].

\bibitem{Garny:2012it}
M.~Garny, A.~Ibarra, M.~Pato, and S.~Vogl, {\it {On the spin-dependent
  sensitivity of XENON100}},  {\em Phys. Rev.} {\bf D87} (2013), no.~5 056002,
  [\href{http://xxx.lanl.gov/abs/1211.4573}{{\tt 1211.4573}}].

\bibitem{PropScintillation}
A.~Lansiart {\em et.~al.}, {\it Development research on a highly luminous
  condensed xenon scintillator},  {\em Nuclear Instruments and Methods} {\bf
  135} (1976), no.~1 47.

\bibitem{NEST_2011}
M.~Szydagis {\em et.~al.}, {\it {NEST}: a comprehensive model for scintillation
  yield in liquid xenon},  {\em Journal of Instrumentation} {\bf 6} (oct, 2011)
  P10002--P10002.

\bibitem{NEST_2013}
M.~Szydagis {\em et.~al.}, {\it Enhancement of {NEST} capabilities for
  simulating low-energy recoils in liquid xenon},  {\em Journal of
  Instrumentation} {\bf 8} (oct, 2013) C10003--C10003.

\bibitem{NEST_2015}
B.~Lenardo {\em et.~al.}, {\it A global analysis of light and charge yields in
  liquid xenon},  {\em IEEE Transactions on Nuclear Science} {\bf 62} (Dec,
  2015) 3387--3396.

\bibitem{FanoFactor_1995}
J.~Seguinot, J.~Tischhauser, and T.~Ypsilantis, {\it Liquid xenon
  scintillation: photon yield and fano factor measurements},  {\em Nuclear
  Instruments and Methods in Physics Research Section A: Accelerators,
  Spectrometers, Detectors and Associated Equipment} {\bf 354} (1995), no.~2
  280 -- 287.

\bibitem{ThomasImel}
J.~Thomas and D.~A. Imel, {\it Recombination of electron-ion pairs in liquid
  argon and liquid xenon},  {\em Phys. Rev. A} {\bf 36} (Jul, 1987) 614--616.

\bibitem{NESTv2}
``Nest v2.'' \url{https://github.com/NESTCollaboration/nest}.

\bibitem{xe1t_physicsreach}
E.~Aprile {\em et.~al.}, {\it Physics reach of the {XENON}1t dark matter
  experiment.},  {\em Journal of Cosmology and Astroparticle Physics} {\bf
  2016} (apr, 2016) 027--027.

\bibitem{xe100_SI}
{\bf XENON100} Collaboration, E.~Aprile {\em et.~al.}, {\it {Dark Matter
  Results from 225 Live Days of XENON100 Data}},  {\em Phys. Rev. Lett.} {\bf
  109} (2012) 181301, [\href{http://xxx.lanl.gov/abs/1207.5988}{{\tt
  1207.5988}}].

\bibitem{zeplin}
G.~Alner {\em et.~al.}, {\it Limits on spin-dependent wimp-nucleon
  cross-sections from the first zeplin-ii data},  {\em Physics Letters B} {\bf
  653} (2007), no.~2 161 -- 166.

\bibitem{lz_proj}
{\bf LUX-ZEPLIN} Collaboration, D.~S. Akerib {\em et.~al.}, {\it {Projected
  WIMP Sensitivity of the LUX-ZEPLIN (LZ) Dark Matter Experiment}},
  \href{http://xxx.lanl.gov/abs/1802.06039}{{\tt 1802.06039}}.

\bibitem{xe1t_SD}
{\bf XENON} Collaboration, E.~Aprile {\em et.~al.}, {\it {Constraining the
  spin-dependent WIMP-nucleon cross sections with XENON1T}},  {\em Phys. Rev.
  Lett.} {\bf 122} (2019), no.~14 141301,
  [\href{http://xxx.lanl.gov/abs/1902.03234}{{\tt 1902.03234}}].

\bibitem{xe1t_analysis}
E.~Aprile {\em et.~al.}, {\it {XENON1T Dark Matter Data Analysis: Signal
  Reconstruction, Calibration and Event Selection}},
  \href{http://xxx.lanl.gov/abs/1906.04717}{{\tt 1906.04717}}.

\bibitem{LUX_analysis}
{\bf LUX Collaboration} Collaboration, D.~S. Akerib {\em et.~al.}, {\it
  Calibration, event reconstruction, data analysis, and limit calculation for
  the lux dark matter experiment},  {\em Phys. Rev. D} {\bf 97} (May, 2018)
  102008.

\bibitem{PandaX_Commissioning}
{\bf PandaX-II Collaboration} Collaboration, A.~Tan {\em et.~al.}, {\it Dark
  matter search results from the commissioning run of pandax-ii},  {\em Phys.
  Rev. D} {\bf 93} (Jun, 2016) 122009.

\bibitem{xe1t_PMTs}
E.~Aprile {\em et.~al.}, {\it Lowering the radioactivity of the photomultiplier
  tubes for the xenon1t dark matter experiment},  {\em The European Physical
  Journal C} {\bf 75} (Nov, 2015) 546.

\bibitem{lux_PMTs}
D.~Akerib {\em et.~al.}, {\it An ultra-low background pmt for liquid xenon
  detectors},  {\em Nuclear Instruments and Methods in Physics Research Section
  A: Accelerators, Spectrometers, Detectors and Associated Equipment} {\bf 703}
  (2013) 1 -- 6.

\bibitem{xe1t_distilllation}
{\bf XENON} Collaboration, E.~Aprile {\em et.~al.}, {\it Removing krypton from
  xenon by cryogenic distillation to the ppq level},  {\em The European
  Physical Journal C} {\bf 77} (May, 2017) 275.

\bibitem{xe1t_screening}
{\bf XENON} Collaboration, E.~Aprile {\em et.~al.}, {\it {Material radioassay
  and selection for the XENON1T dark matter experiment}},  {\em Eur. Phys. J.}
  {\bf C77} (2017), no.~12 890, [\href{http://xxx.lanl.gov/abs/1705.01828}{{\tt
  1705.01828}}].

\bibitem{pandaX_screening}
X.~Wang, X.~Chen, C.~Fu, X.~Ji, X.~Liu, Y.~Mao, H.~Wang, S.~Wang, P.~Xie, and
  T.~Zhang, {\it Material screening with {HPGe} counting station for {PandaX}
  experiment},  {\em Journal of Instrumentation} {\bf 11} (dec, 2016)
  T12002--T12002.

\bibitem{xe100_rnremoval}
{\bf XENON100} Collaboration, E.~Aprile {\em et.~al.}, {\it {Online$^{222}$ Rn
  removal by cryogenic distillation in the XENON100 experiment}},  {\em Eur.
  Phys. J.} {\bf C77} (2017), no.~6 358,
  [\href{http://xxx.lanl.gov/abs/1702.06942}{{\tt 1702.06942}}].

\bibitem{hexe_distillation}
S.~Bruenner, D.~Cichon, S.~Lindemann, T.~Marrodan~Undagoitia, and H.~Simgen,
  {\it {Radon depletion in xenon boil-off gas}},  {\em Eur. Phys. J.} {\bf C77}
  (2017), no.~3 143, [\href{http://xxx.lanl.gov/abs/1611.03737}{{\tt
  1611.03737}}].

\bibitem{lz_adsorption}
K.~Pushkin {\em et.~al.}, {\it Study of radon reduction in gases for rare event
  search experiments},  {\em Nuclear Instruments and Methods in Physics
  Research Section A: Accelerators, Spectrometers, Detectors and Associated
  Equipment} {\bf 903} (2018) 267 -- 276.

\bibitem{exo_xe136}
{\bf EXO-200} Collaboration, J.~B. Albert {\em et.~al.}, {\it {Improved
  measurement of the $2\nu\beta\beta$ half-life of $^{136}$Xe with the EXO-200
  detector}},  {\em Phys. Rev.} {\bf C89} (2014), no.~1 015502,
  [\href{http://xxx.lanl.gov/abs/1306.6106}{{\tt 1306.6106}}].

\bibitem{LZ}
B.~J. Mount {\em et.~al.}, {\it {LUX-ZEPLIN (LZ) Technical Design Report}},
  \href{http://xxx.lanl.gov/abs/1703.09144}{{\tt 1703.09144}}.

\bibitem{Billard_neutrinofloor}
J.~Billard, L.~Strigari, and E.~Figueroa-Feliciano, {\it {Implication of
  neutrino backgrounds on the reach of next generation dark matter direct
  detection experiments}},  {\em Phys. Rev.} {\bf D89} (2014), no.~2 023524,
  [\href{http://xxx.lanl.gov/abs/1307.5458}{{\tt 1307.5458}}].

\bibitem{Pirinen:2018gsd}
P.~Pirinen, J.~Suhonen, and E.~Ydrefors, {\it {Neutral-current neutrino-nucleus
  scattering off Xe isotopes}},  \href{http://xxx.lanl.gov/abs/1804.08995}{{\tt
  1804.08995}}.

\bibitem{Cowan:2010js}
G.~Cowan, K.~Cranmer, E.~Gross, and O.~Vitells, {\it {Asymptotic formulae for
  likelihood-based tests of new physics}},  {\em Eur. Phys. J.} {\bf C71}
  (2011) 1554, [\href{http://xxx.lanl.gov/abs/1007.1727}{{\tt 1007.1727}}].
  [Erratum: Eur. Phys. J.C73,2501(2013)].

\bibitem{Agashe:2014kda}
{\bf Particle Data Group} Collaboration, K.~A. Olive {\em et.~al.}, {\it
  {Review of Particle Physics}},  {\em Chin. Phys.} {\bf C38} (2014) 090001.

\bibitem{Wilks:1938dza}
S.~S. Wilks, {\it {The Large-Sample Distribution of the Likelihood Ratio for
  Testing Composite Hypotheses}},  {\em Annals Math. Statist.} {\bf 9} (1938),
  no.~1 60--62.

\bibitem{10.2307/1990256}
A.~Wald, {\it Tests of statistical hypotheses concerning several parameters
  when the number of observations is large},  {\em Transactions of the American
  Mathematical Society} {\bf 54} (1943), no.~3 426--482.

\bibitem{Aprile:2019dbj}
E.~Aprile {\em et.~al.}, {\it {Constraining the spin-dependent WIMP-nucleon
  cross sections with XENON1T}},
  \href{http://xxx.lanl.gov/abs/1902.03234}{{\tt 1902.03234}}.

\bibitem{Rogers:2016jrx}
H.~Rogers, D.~G. Cerdeno, P.~Cushman, F.~Livet, and V.~Mandic, {\it
  {Multidimensional effective field theory analysis for direct detection of
  dark matter}},  {\em Phys. Rev.} {\bf D95} (2017), no.~8 082003,
  [\href{http://xxx.lanl.gov/abs/1612.09038}{{\tt 1612.09038}}].

\bibitem{Fieguth:2018vob}
A.~Fieguth, M.~Hoferichter, P.~Klos, J.~Menendez, A.~Schwenk, and
  C.~Weinheimer, {\it {Discriminating WIMP-nucleus response functions in
  present and future XENON-like direct detection experiments}},
  \href{http://xxx.lanl.gov/abs/1802.04294}{{\tt 1802.04294}}.

\bibitem{Catena:2017wzu}
R.~Catena, J.~Conrad, C.~Doring, A.~D. Ferella, and M.~B. Krauss, {\it {Dark
  matter spin determination with directional direct detection experiments}},
  {\em Phys. Rev.} {\bf D97} (2018), no.~2 023007,
  [\href{http://xxx.lanl.gov/abs/1706.09471}{{\tt 1706.09471}}].

\bibitem{Kahlhoefer:2016eds}
F.~Kahlhoefer and S.~Wild, {\it {Studying generalised dark matter interactions
  with extended halo-independent methods}},  {\em JCAP} {\bf 1610} (2016),
  no.~10 032, [\href{http://xxx.lanl.gov/abs/1607.04418}{{\tt 1607.04418}}].

\end{thebibliography}\endgroup

\clearpage

\end{document}